\documentclass[namedreferences]{solarphysics}
\usepackage[optionalrh,solaromanenum]{spr-sola-addons} 
\usepackage{graphicx}                    
\usepackage{color}                       
\usepackage{url}                         

\begin{document}

\begin{article}

\begin{opening}

\title{Bridging EUV and white-light observations to inspect the initiation phase of a ``two-stage" solar eruptive event.}

%
\author{J.\,P.~\surname{Byrne}$^{1}$\sep
	H.~\surname{Morgan}$^{2}$\sep
	D.\,B.~\surname{Seaton}$^{3}$\sep
	H.\,M.~\surname{Bain}$^{4}$\sep     
	S.\,R.~\surname{Habbal}$^{1}$
       }

%
\runningauthor{Byrne et al.}
\runningtitle{Bridging EUV and white-light observations of a ``two-stage" solar eruptive event.}

%
  \institute{$^{1}$Institute for Astronomy, University of Hawai'i, 2680 Woodlawn Drive, Honolulu, HI 96822, USA.
                     email: \url{jbyrne@ifa.hawaii.edu} 
	$^{2}$Sefydliad Mathemateg a Ffiseg, Prifysgol Aberystwyth, Ceredigion, Cymru, SY23 3BZ, UK.
	$^{3}$Royal Observatory of Belgium, Avenue Circulaire 3, 1180 Brussels, Belgium.
	$^{4}$Space Sciences Laboratory, University of California, Berkeley, CA 94720-7450, USA.
             }

\begin{abstract}
The initiation phase of CMEs is a very important aspect of solar physics, as these phenomena ultimately drive space weather in the heliosphere. This phase is known to occur between the photosphere and low corona, where many models introduce an instability and/or magnetic reconnection that triggers a CME, often with associated flaring activity. To this end, it is important to obtain a variety of observations of the low corona in order to build as clear a picture as possible of the dynamics that occur therein. Here, we combine the EUV imagery of the SWAP instrument on board \emph{PROBA2} with the white-light imagery of the ground-based Mk4 coronameter at MLSO in order to bridge the observational gap that exists between the disk imagery of AIA on board \emph{SDO} and the coronal imagery of LASCO on board \emph{SOHO}. Methods of multiscale image analysis were applied to the observations to better reveal the coronal signal while suppressing noise and other features. This allowed an investigation into the initiation phase of a CME that was driven by a rising flux rope structure from a ``two-stage" flaring active region underlying an extended helmet streamer. It was found that the initial outward motion of the erupting loop system in the EUV observations coincided with the first X-ray flare peak, and led to a plasma pile-up of the white-light CME core material. The characterized CME core then underwent a strong jerk in its motion, as the early acceleration increased abruptly, simultaneous with the second X-ray flare peak. The overall system expanded into the helmet streamer to become the larger CME structure observed in the LASCO coronagraph images, which later became concave-outward in shape. Theoretical models for the event are discussed in light of these unique observations, and it is concluded that the formation of either a kink-unstable or torus-unstable flux rope may be the likeliest scenario.
 
\end{abstract}

%

\end{opening}

%

\section{Introduction}
\label{intro}

Coronal mass ejections (CMEs) represent the largest, most dynamic phenomena that originate from the Sun \cite{2012LRSP....9....3W}. Propagating at speeds of hundreds up to thousands of kilometers per second \cite{2004JGRA..10907105Y}, the particle densities and energies involved can cause shocks and adverse space weather at Earth and elsewhere in the heliosphere \cite{2004Natur.432...78P,2005AnGeo..23.1033S,2013NatPh...9..811C,2014NatCo...5E3481L}. They can lead to geomagnetic storms upon impacting our magnetosphere, damaging satellites, affecting communication and navigation systems, and increasing the radiation risk for astronauts \cite{2007A&G....48f..11L}. Given their potentially hazardous impact on Earth's geomagnetic environment, the physics governing their eruption and propagation needs to be understood. 

Various theoretical CME models exist in the effort to reproduce the physical driver mechanisms responsible for their initiation and propagation (see the review by \opencite{2011LRSP....8....1C}). All are based upon some form of triggering mechanism, most likely as a result of a magnetic energy imbalance often described in the context of tether straining and release \cite{2001AGUGM.125..143K}. Considering the pre-eruption structure of the CME as a flux rope (e.g., \opencite{1996JGR...10127499C}) or strongly sheared arcade, possible causes for eruption may include flux injection and magnetic twisting \cite{2001ApJ...562.1045K,2006PhRvL..96y5002K}, reconnection beneath the flux rope \cite{1980IAUS...91..207M,1995ApJ...446..377F,2003ApJ...595.1231A,2007ApJ...658L.123L}, or reconnection between the overlying field and neighboring flux systems \cite{1999ApJ...510..485A,2007ApJ...671L..77V,2008ApJ...683.1192L}. Such models may provide an interpretation on observations, and thus allow some deeper understanding of the forces governing CMEs and their relationship with associated phenomena like flares (see the review by \opencite{2002A&ARv..10..313P}).

An important aspect of studying CME initiation, is the ability to resolve their low-coronal propagation and associated source regions on the disk: be it a flaring or non-flaring active region, a prominence/filament eruption or other rising loop system \cite{2001ApJ...561..372S,2002ApJ...566L.117Z}, or else a ``stealth CME" without any specifically detectable source \cite{2013SoPh..285..269H}. Prominence lift-offs often become the core material of a CME \cite{2003ApJ...586..562G,2008AnGeo..26.3025F}, and rising loops often form some part of the CME morphology \cite{2004A&A...422..307C,2006A&A...455..339D}.

The low-coronal kinematics and morphology of CMEs provide insight into the early forces at play, and so a rigorous study of such phenomena is key to understanding the physics involved. However, a difficulty exists in studies of coronal structure that are prone to low signal-to-noise ratios in the observational data. Low-coronal white-light observations using a coronagraph are problematic due to the strong radial brightness gradient and issues with scattered light in the instrument, while extended EUV disk observations are problematic due to the strong drop-off in emission brightness with increasing coronal height. These common issues with solar observational data motivate the ongoing development and use of advanced image processing techniques to suppress noise and enhance structures in the image data \cite{2003A&A...398.1185S,2006SoPh..236..263M,2008SoPh..248..457Y,2011igi-global,2011ApJ...737...88D}. 

In this paper, a relatively unique ``two-stage" solar eruptive event is studied with a combination of multiple, overlapping EUV disk observations and white-light coronal observations. This compliments the study of \inlinecite{2012ApJ...746L...5S}, who reported this event as evidence for secondary heating during a flare and associated CME. In Section\,\ref{sect:techniques} we describe the observations and use of multiscale image processing methods. In Section\,\ref{sect:event} we describe the event, that occurred on 8~March~2011, and how the combination of observations and techniques can provide deeper insight to the initiation phase of CMEs. A discussion of the interpretation of this study is presented in Section\,\ref{sect:discussion}, and final conclusions in Section\,\ref{sect:conclusions}.

\section{Observations \& Techniques}
\label{sect:techniques}

\begin{figure}[ht]
\centering{\includegraphics[scale=0.6, trim=60 50 200 70]{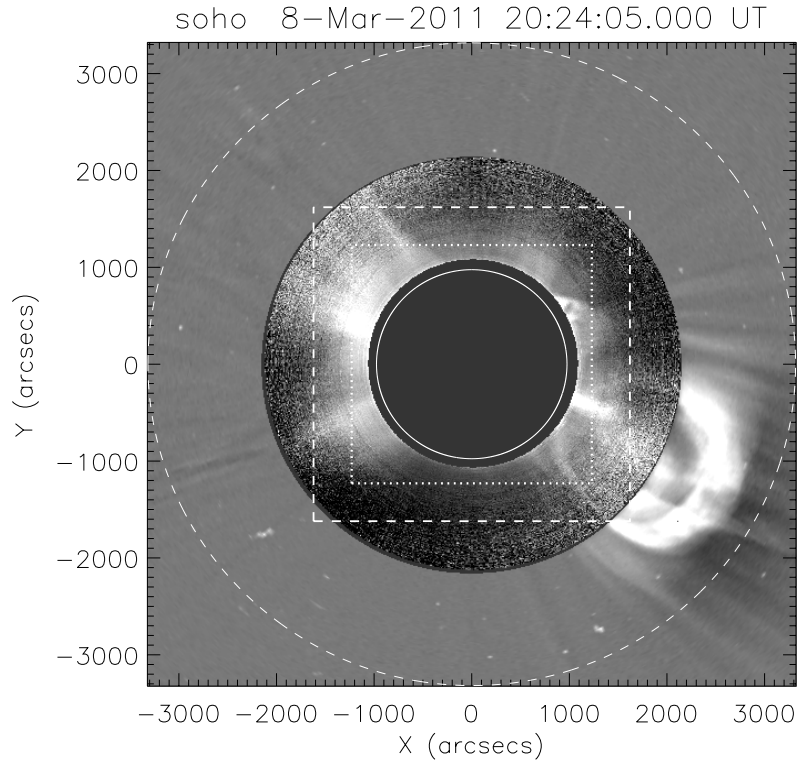}}
\caption{A zoomed-in LASCO/C2 image of the 8~March~2011 CME, with an MLSO/Mk4 image overlaid in the range 1.1\,--\,2.2\,$R_\odot$, at times 20:24 and 20:22\,UT respectively. The C2 image has been processed via the CORIMP techniques of normalizing radial graded filter (NRGF) and quiescent background subtraction. It has been trimmed to a half-width of 3.4\,$R_\odot$, which is the upper limit of the \emph{PROBA2}/SWAP field-of-view as indicated by the dashed circle. The SWAP field-of-view during nominal operations is indicated by the outer dashed box, and the \emph{SDO}/AIA field-of-view is indicated by the inner dotted box. The limb of the Sun behind the occulter is indicated by the solid white circle. A CME is observed off the south-west limb as a bright loop structure with some inner core material, as seen here in the Thomson-scattered white-light coronagraph images. It is clear how the fields-of-view of SWAP and AIA can be used to bridge CME observations to the low corona and solar disk, for gaining insight to their initiation phase.}
\label{overlays}
\end{figure}

In order to connect CMEs to their source regions, data from disk imagers such as the Sun Watcher using APS detectors and image Processing (SWAP; \opencite{2013SoPh..286...43S}, \opencite{2013SoPh..286...67H}) on board the second Projects for Onboard Autonomy (\emph{PROBA2}; \opencite{2013SoPh..286....5S}) and the Atmospheric Imaging Assembly (AIA; \opencite{2012SoPh..275...17L}) on board the Solar Dynamics Observatory (\emph{SDO}; \opencite{2012SoPh..275....3P}), may be used in tandem with coronagraph observations. However, difficulties in the interpretation of the observed features arise due to the varying instrument specifications, particularly the limitations on their fields-of-view, image passbands, and cadences of observations. In order to bridge the gap between the EUV observations of the low corona and the white-light images of the upper corona, the SWAP imager was used in conjunction with the ground-based Mauna Loa Solar Observatory (MLSO) Mk4 coronameter \cite{2003SPIE.4843...66E} to directly compare the observations of CMEs as they erupt through the overlapping fields-of-view (Fig.\,\ref{overlays}). SWAP has a spectral bandpass centered on 174\,{\AA}, with 3.2 arcsec pixels over a 54$\times$54 arcmin field-of-view, and a cadence of $\sim$\,2~minutes. Mk4 is a rotating coronameter that produces white-light images of the polarization brightness of the corona from 1.1\,--\,2.8\,$R_{\odot}$ at a cadence of $\sim$\,3~minutes over a five-hour observing day. This allows a direct correspondence of features in the EUV images with those in the white-light images, and can therefore provide new insight into the initial phases of CME eruption and acceleration.

\begin{figure}[t]
\centering{\includegraphics[width=\linewidth]{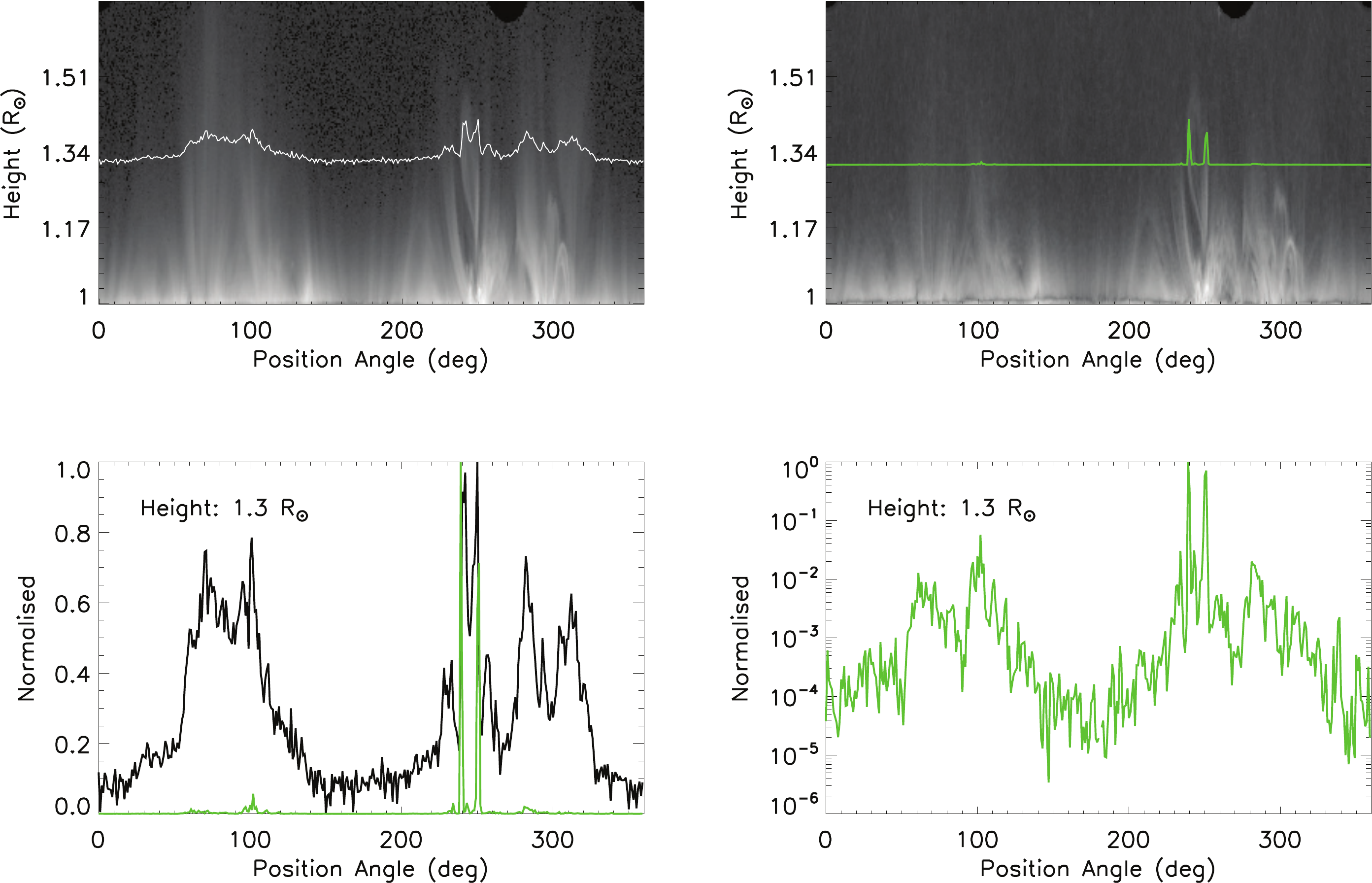}}
\caption{The top two panels show polar-unwrapped images of the solar corona across the \emph{PROBA2}/SWAP field-of-view on 8~March~2011 at 19:53:59\,UT; left being the level-1 data, right being the multiscale enhanced data (shown at log-scale image intensity). Across each image, at a constant height of 1.3\,$R_{\odot}$, a normalized intensity profile is plotted to demonstrate how the background coronal structure is suppressed by the multiscale techniques, to highlight only the complex structure of the prominence. The bottom left plot shows a direct comparison of the two intensity profiles, where the prominence is located between 230\,--\,260$^\circ$. The bottom right plot shows a log scale of the normalized intensity profile across the enhanced image to demonstrate that the rest of the coronal structure is still present, just strongly suppressed relative to the prominence material. (The green line in each case corresponds to the intensity slice of the multiscale enhanced image at top right.)}
\label{polar_fig_swap_20110308}
\end{figure}

Methods of multiscale image processing have been developed in recent years for use on coronagraph images to enhance the underlying structure \cite{2008ApJ...674.1201S,2008SoPh..248..457Y,2011AdSpR..47.2118G}. The fundamental idea behind these methods is to highlight details apparent on different scales within the data. Therefore, multiscale techniques provide an ability to remove small-scale features in images, essentially suppressing the noise such that the structures of interest can be revealed in greater detail. By applying them to coronagraph images, the morphology of CMEs as they propagate through the corona in a sequence of observations can be determined with better accuracy than previously possible, and can allow a characterization of the erupting structure to determine various properties in their evolution (\opencite{2009A&A...495..325B}, \citeyear{2010NatCo...1E..74B}, \citeyear{2012ApJ...752..145B}). 

Here, multiscale methods are demonstrated for use on the Mk4 coronameter and SWAP EUV imager, to provide insight to the low-coronal morphology of erupting structures that form CMEs. Details on the fundamental techniques are outlined in \inlinecite{2008SoPh..248..457Y} wherein the magnitude of the multiscale gradient is used to show the relative strength of the detected edges in the image structure at a particular scale of the multiscale decomposition (i.e., the strongest edges appear brightest). To further increase the signal-to-noise ratio of the edge detections, the magnitude information from the scales most relevant to the coronal structures of interest may be multiplied together, neglecting the largest scales that smooth out the coronal signal, and the smallest scales that reveal the finer structure and noise (see \opencite{2012ApJ...752..145B}, for details). Thus the magnitude of the multiscale gradient across the dominant edges of coronal loops and CMEs is further enhanced for subsequent characterization of their morphology. 

Figure\,\ref{polar_fig_swap_20110308} shows the effectiveness of the multiscale techniques for detecting the structure of an ejection observed by SWAP at 19:53:59\,UT on 8~March~2011. The top left image shows the level-1 processed data, polar-unwrapped about Sun-center at coronal heights of 1\,--\,1.7\,$R_{\odot}$. The top right image shows the result of the multiscale filtering technique applied in such a manner as to enhance the edges of the detected structure in the data. The bottom left plot shows the comparison of a normalized intensity profile at a height of 1.3\,$R_{\odot}$ in each, revealing how the multiscale techniques best characterize the complex structure of the erupting prominence material located between  230\,--\,260$^\circ$. The bottom right plot shows a log scale of the normalized intensity profile across the multiscale filtered image to reveal the suppressed signal of the background features.
\begin{figure}[p]
\centering{\includegraphics[scale=0.21, clip=true]{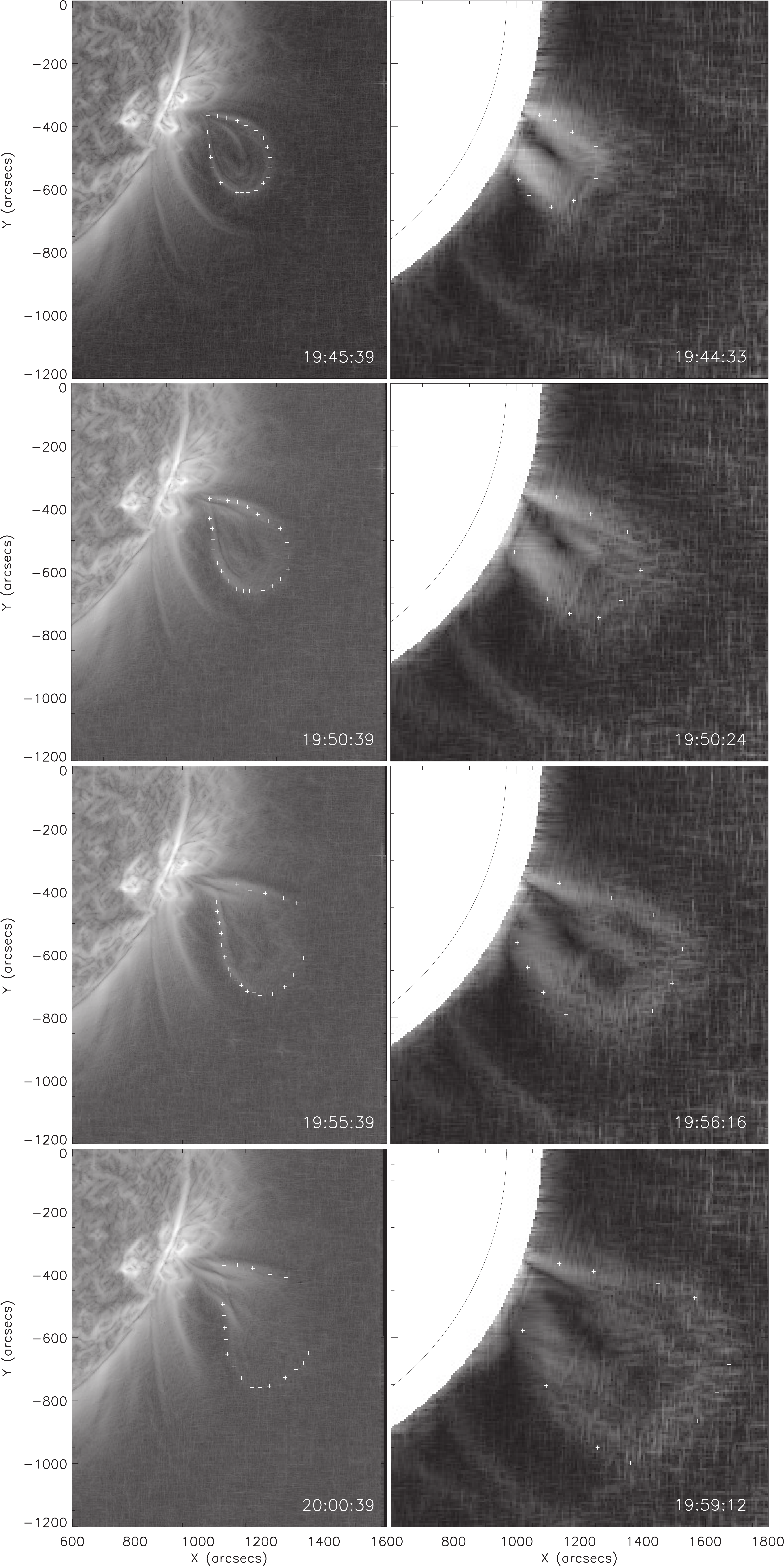}}
\caption{SWAP (left) and Mk4 (right) observations of the erupting loop system that formed the core of the CME on 8~March~2011. The images have been processed using a multiscale filter, and have intensities that represent the relative magnitudes of the edges in each image. The images are zoomed to the same size scale, with Mk4 extending further in the x-direction. The erupting structure was traced manually by a point-\&-click characterization, as shown.}
\label{combined_modgrad_points}
\end{figure}

The extended corona from $\sim$\,2\,--\,30\,$R_{\odot}$ is often observed with the Large Angle Spectrometric Coronagraph (LASCO; \opencite{1995SoPh..162..357B})  on board the Solar and Heliospheric Observatory (SOHO; \opencite{1995SoPh..162....1D}) which orbits the L1 point. Coronal structures, and specifically CMEs, have been studied in the white-light image data from these instruments through the use of a number of steps outlined in the Coronal Image Processing package (CORIMP; \opencite{2012ApJ...752..144M}; \opencite{2012ApJ...752..145B}). Here, these techniques are extended for use on the MLSO/Mk4 coronameter data. The Mk4 data is prepared via an instrumental vignetting function that maximizes the image contrast by offsetting the radial brightness gradient in order to best reveal structures such as CMEs and streamers. The multiscale methods are then applied to the data in order to produce magnitude images of the relative edge strengths in the images, and highlight the detected structure. Figure\,\ref{combined_modgrad_points} shows four multiscale enhanced images from each of the SWAP and Mk4 observations of the 8~March~2011 event, allowing a point-\&-click characterization of the erupting structure of interest. This allows, for example, an ellipse-fit to the outward propagating fronts, to obtain kinematical and morphological information as described in \inlinecite{2009A&A...495..325B}.

\section{``Two-Stage" Solar Eruptive Event}
\label{sect:event}

\begin{figure}[t]
\centering{\includegraphics[width=\linewidth, clip=true, trim=0 0 120 150]{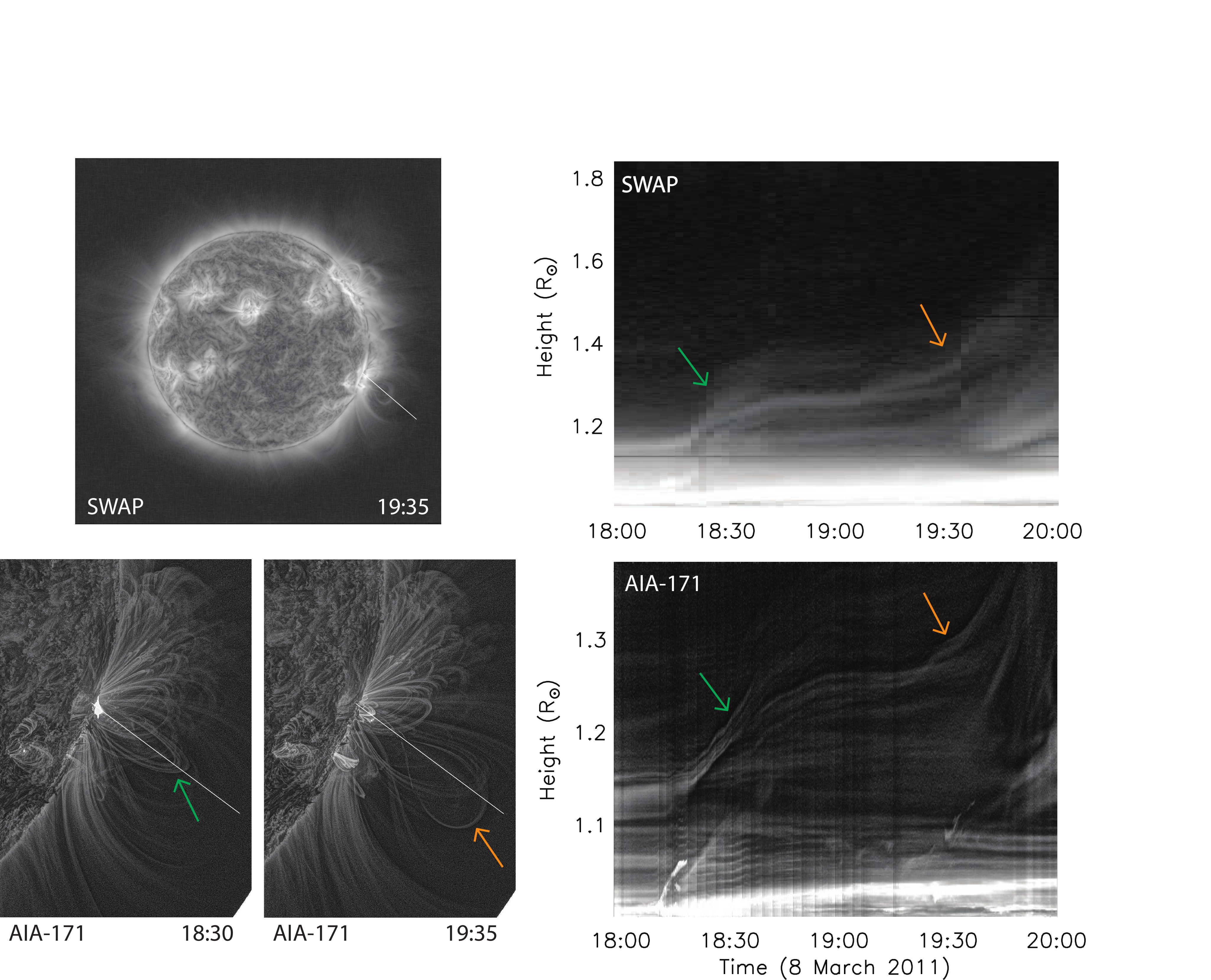}}
\caption{Stack plots were generated for a radial intensity profile through the enhanced SWAP and AIA-171 observations. The profile location is shown in a full field-of-view SWAP image at 19:35\,UT (top left), and at two stages of the eruption in cropped AIA-171 images at 18:30 and 19:35\,UT on 8~March~2011 (bottom left). The stack plots (right) clearly show the two-stage eruption, with green and orange arrows to indicate the different loop structures of interest; the latter of which becomes the CME core that is tracked through the SWAP, Mk4 and LASCO observations. (An online movie of the enhanced AIA-171 observations accompanies this figure.)}
\label{stackplots}
\end{figure}

A CME erupted from the southwest active region NOAA\,11165, first visible in LASCO/C2 at 20:12\,UT on 8~March~2011. The active region caused numerous flares during its evolution across the disk, notably an M4.4 flare at GOES start-time 18:08\,UT (peaking at 18:28\,UT) associated with the rising loop system that later erupted to form the core material of the CME. Of particular interest for this event, is the ``two-stage" X-ray flare profile seen in the GOES flux, identified as two individual M-class flares separated by almost 2 hours. \inlinecite{2012ApJ...746L...5S} present this as evidence for a secondary heating phase. The loop system evolution was visible up to $\sim$\,1.3\,$R_{\odot}$ in AIA images, at which height a set of loops that were most strongly observed in AIA\,171\,{\AA} images began to erupt, coinciding with the time of the secondary M1.4 flare $\sim$\,20:00\,UT. These were observable to a height of $\sim$\,1.6\,$R_{\odot}$ in the larger field-of-view of the SWAP\,174\,{\AA} imager. The observations of the source active region are exhaustively reported by \inlinecite{2012ApJ...746L...5S}, though it is noted that the system of loops they track (in the stack plot of Fig.\,4 of their paper) is not the same as the specific flux rope loops observed and tracked here, which erupted as the core of the CME. Figure\,\ref{stackplots} shows these different sets of loops that correspond to the two stages of the eruption. The CME core was then observed in the white-light Mk4 coronameter images to a height of $\sim$\,2.2\,$R_{\odot}$, entrained within a faint CME bubble that started to become visible at these heights before being clearly observed to propagate outwards in the extended LASCO coronagraph images. This type of cavity-CME morphology is often seen in low coronal observations, studied for example by \inlinecite{2006ApJ...641..590G} for long-lasting structures seen in Mk4 data, which erupt to become the typical three-part CME structure of a bright front, darker cavity and bright core \cite{1986JGR....9110951I}.  

\begin{figure}[!t]
\centering{\includegraphics[width=\linewidth, clip=true, trim=30 300 30 30]{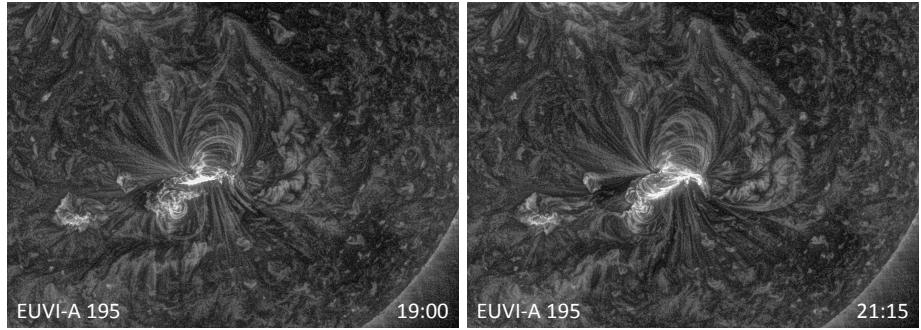}}
\caption{EUVI-A 195\,{\AA} observations of the active region NOAA\,11165 at times 19:00 and 21:15\,UT on 8~March~2011. The left image shows the post-flare loops of the first stage of the eruption, following the initial X-ray flare peak at $\sim$18:28\,UT; while the right image shows the post-flare loops of the second stage of the eruption, following the secondary X-ray flare peak at $\sim$20:15\,UT. (An online movie accompanies this figure.)}
\label{euvi}
\end{figure}

The Solar Terrestrial Relations Observatory (\emph{STEREO}; \opencite{2008SSRv..136....5K}) Ahead spacecraft was at a separation of almost 88$^{\circ}$ from Earth, providing direct observations of the active region with the Sun-Earth Connection Coronal \& Heliospheric Investigation (SECCHI; \opencite{2008SSRv..136...67H}) Extreme Ultraviolet Imager (EUVI; \opencite{2004SPIE.5171..111W}) at wavelengths of 171\,{\AA}, 195\,{\AA}, 284\,{\AA} and 304\,{\AA}. These images, particularly the 195\,{\AA}, reveal a top-down view of the loop structures involved in the two-stage flaring eruption. Initial post-flare loops were observed following the M-class flare X-ray peak from about 19:00\,UT, proceeded by relatively higher post-flare loops following the secondary X-ray peak from about 21:15\,UT (Fig.\,\ref{euvi}). The outward propagation of the erupting loop structure is also very faintly visible, with a slow and steady rise before the secondary eruption, and appearing to have underwent a slight clockwise untwisting motion and/or possible asymmetric expansion (see AIA movie). Thus it may be that the erupting loops, oriented north-south above the east-west flare ribbons seen in EUVI-Ahead, were impeded by overlying field lines oriented in a more east-west direction as seen in the AIA images.

\begin{figure}[t]
\centering{\includegraphics[width=\linewidth, trim=60 0 0 0]{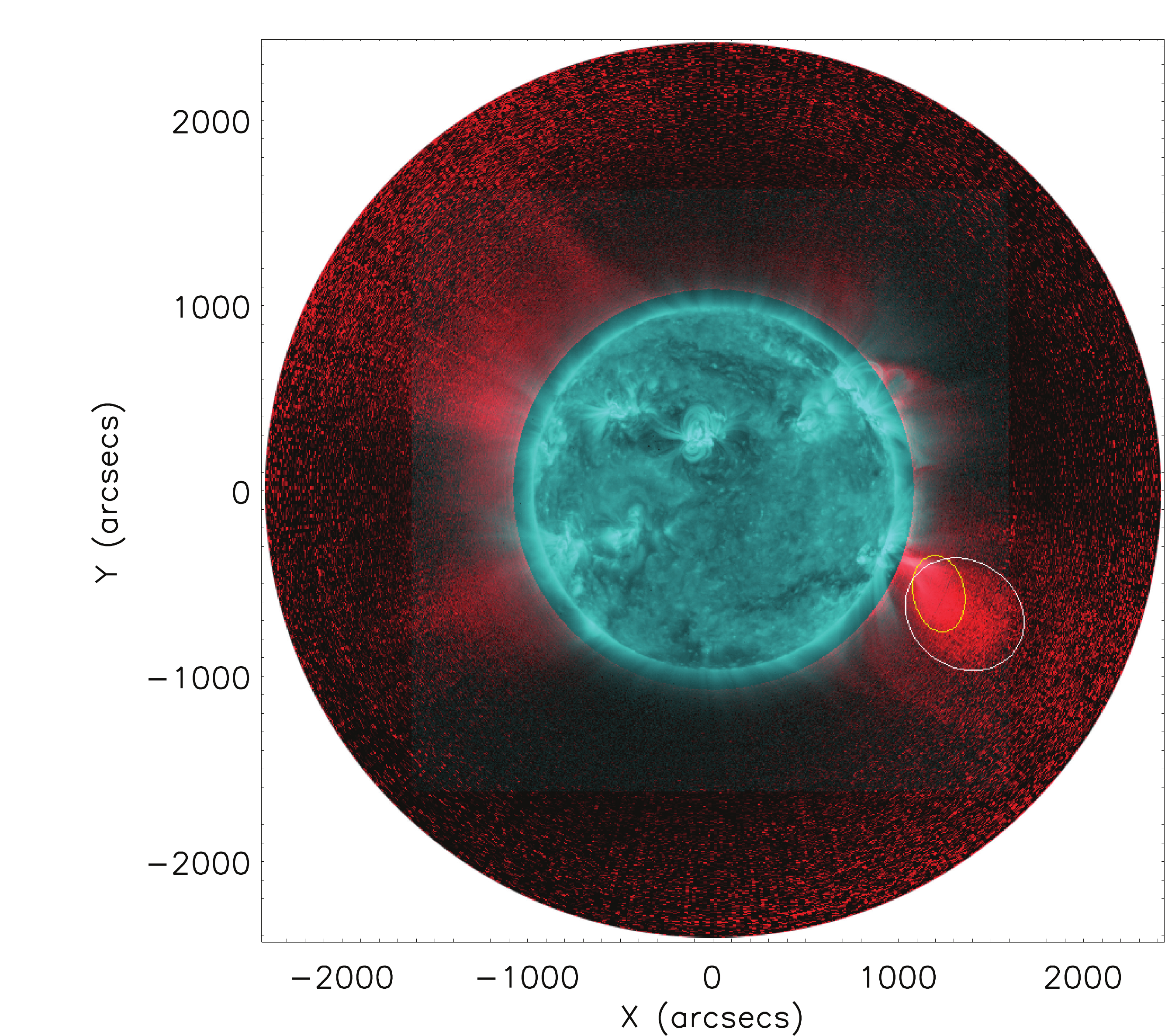}}
\caption{A merged SWAP (blue) and Mk4 (red) image with the ellipse-fits to the characterized CME core material as observed by each instrument at 19:59\,UT on 8 March 2011. (An online movie accompanies this figure.)}
\label{combined}
\end{figure}

\inlinecite{2013ApJS..206...19M} report on the expansion of active region loops from this region into the extended solar corona in the few days leading up to this CME. The region lay beneath a helmet streamer structure that appeared to contain the observed coronal loops, with a number of faint brightenings due to small outward-propagating plasma blobs. These, and the pointed shape of the rising loops, are postulated to be indicators of helmet streamer interchange reconnection at the apex of the closed field \cite{2012ApJ...749..182W}. A subsequent brightening and expanding of the loops, accompanied by a swelling of the helmet streamer, preceded the CME from this region and is evidence for an energy input to the system that leads to an explosive energy release. Such a process manifested as the two-stage solar eruptive event outlined here and in the observations of \inlinecite{2012ApJ...746L...5S}. The AIA images of the active region underlying this system were processed with a multiscale Gaussian normalization technique \cite{MorganDruckmuller_inreview} to enhance the observations of the coronal loop evolution over the course of the event. The right panels of Fig.\,\ref{stackplots} show stack plots obtained from a radial intensity profile through both the enhanced SWAP and AIA-171 observations. The stack plots clearly show the two stages of the eruption, with green and orange arrows on the AIA images indicating the different loop structures that each intensity track corresponds to; the latter being the flux rope structure that erupts as the CME core. The loop structures in the AIA-171 images also exhibit a kink-unstable topology, with a single twist observed most clearly after the first stage of the eruption. This is especially highlighted by the path of upflowing material into the southern portion of the loops, tracing the kinked loop structure (see AIA movie).	 

\begin{figure}[t]
\centering{\includegraphics[scale=0.82, clip=true, trim=10 0 10 60]{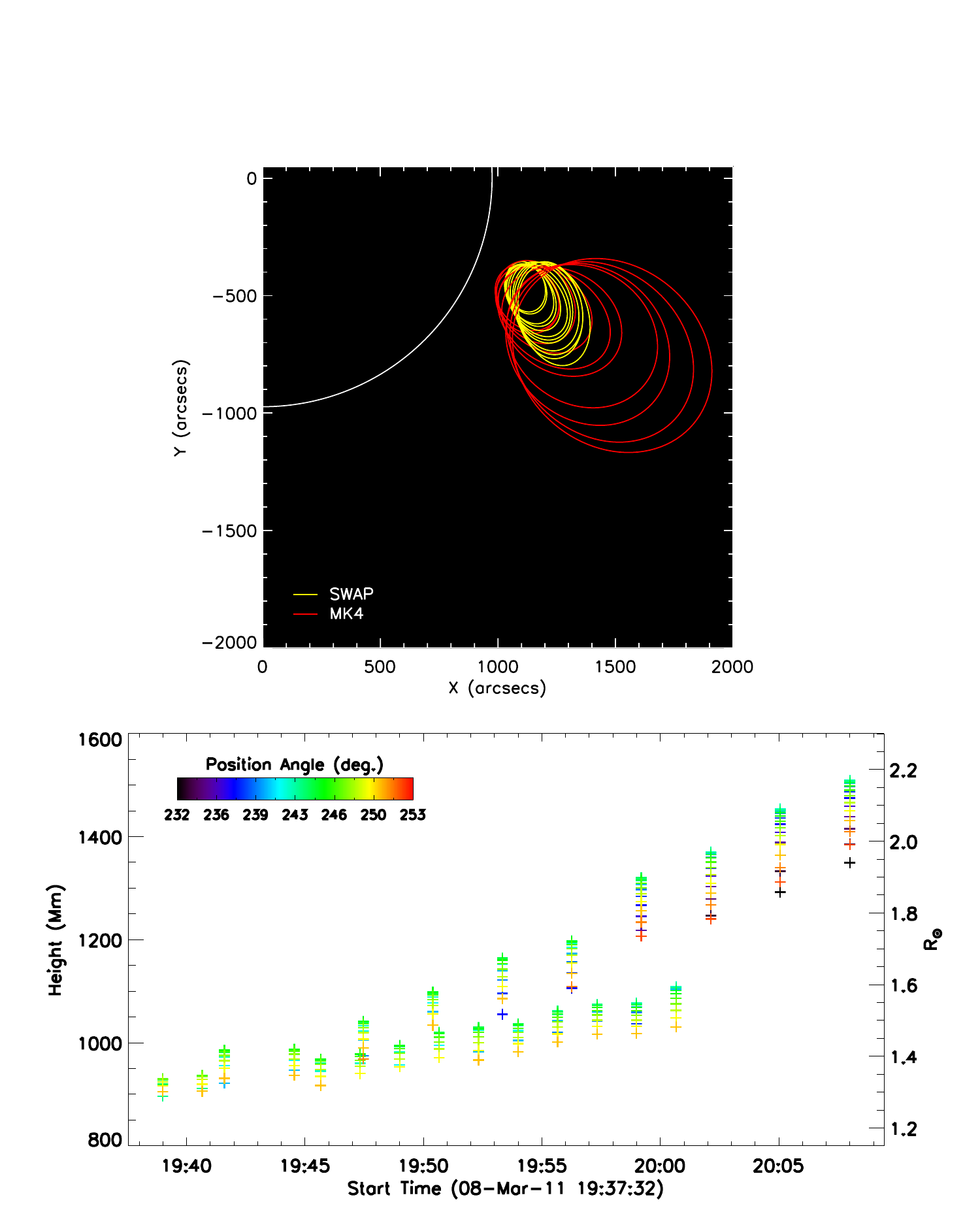}}
\caption{\emph{Top:} The SWAP and Mk4 ellipse-fits to the characterized CME core material over the course of the eruption. \emph{Bottom:} The height-time profile of the characterized eruption observed simultaneously with the SWAP imager and Mk4 coronameter, where the plus symbols represent the span of heights of each of the ellipse-fits, as indicated by the colorbar. The spatial and temporal offsets are due to the different speeds of the eruptions and difference cadences of the instruments; where the erupting EUV loops observed with SWAP moved at a speed of $\sim$\,100$\,km\,s^{-1}$, while the associated core of the CME observed with Mk4 reached a speed of $\sim$\,400$\,km\,s^{-1}$.}
\label{ell_heights_inner}
\end{figure}

In order to best reveal the eruption material when comparing the low signal-to-noise SWAP and Mk4 images, multiscale methods of noise suppression and edge enhancement were employed, as discussed above. This allowed a robust point-\&-click characterization of the CME core material, which was the brightest structure to be tracked through the different imagers when the CME front was not yet fully formed (Fig.\,\ref{combined_modgrad_points}). The rising loop system observed with SWAP and the erupting CME core material observed with Mk4 coincided both temporally and spatially, at least initially, and each was characterized by ellipse-fits to the detected front edges of the core flux rope structure. Figure\,\ref{combined} shows an overlay of SWAP and Mk4 images during the eruption at times 19:58:59 and 19:59:12\,UT respectively, with the ellipse-fits to the erupting fronts. Figure\,\ref{ell_heights_inner} shows the progression of the ellipse-fits over the course of the eruption, indicating how the white-light material observed with Mk4 propagated away from the source quicker than the EUV material observed with SWAP. The different height-time profiles show the material in the Mk4 images attained a speed of $\sim$\,400$km\,s^{-1}$, while the associated erupting loop structures in the SWAP images moved at only $\sim$\,100$\,km\,s^{-1}$. This may be, at least partly, due to the increased intensity drop-off of the EUV emission with height, and possible plasma pile-up in the white-light observations. It is also important to note that Mk4 observes the real density-enhanced structure, while the SWAP EUV is also sensitive to the temperature of the coronal material.

\begin{figure}[t]
\centering{\includegraphics[clip=true, scale=0.4, trim=0 60 0 90]{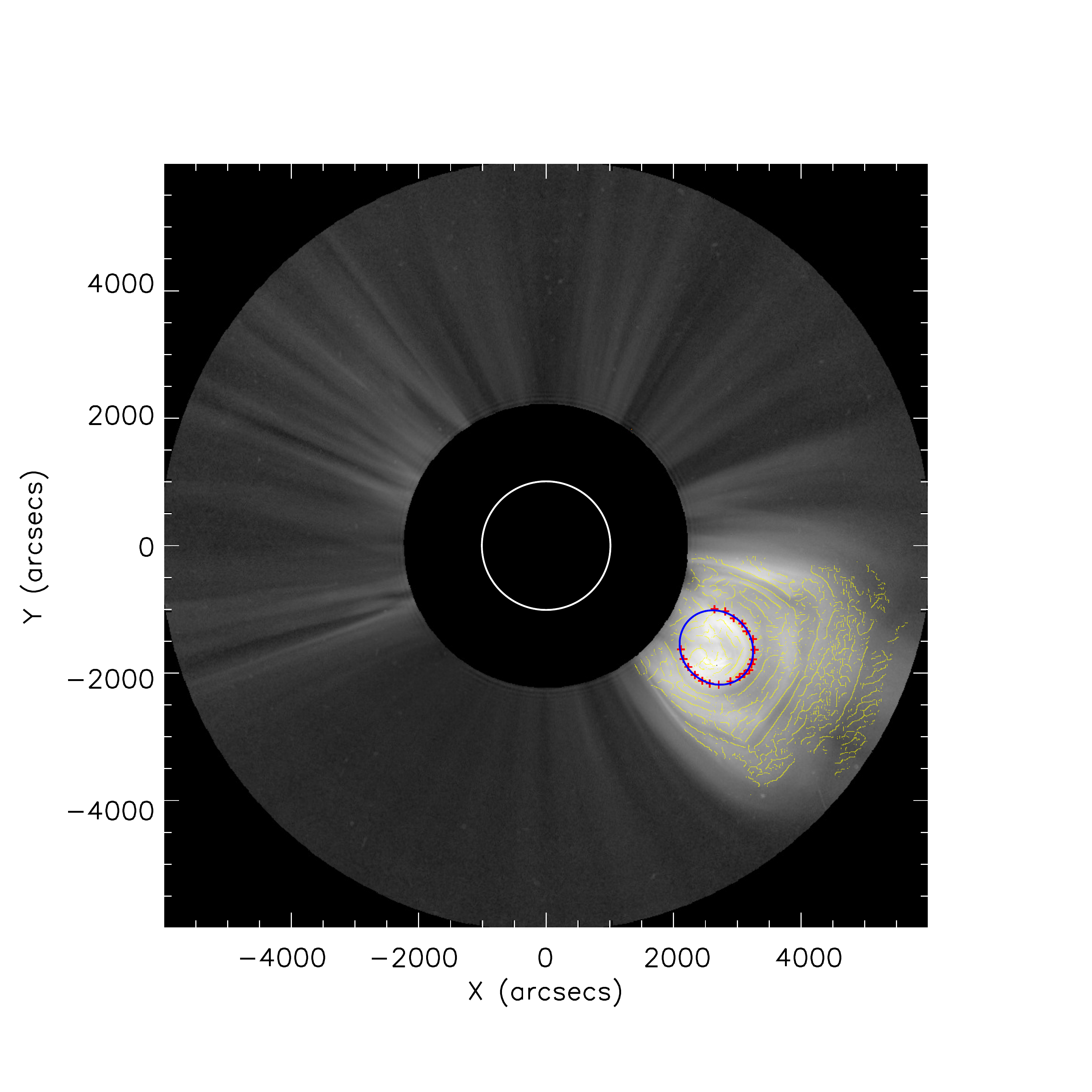}}
\caption{A radially filtered LASCO/C2 image of the CME at 21:06\,UT on 8~March~2011. The yellow contours trace the edges in the detected CME structure (from the automated CORIMP catalog), with red points manually clicked along the corresponding core material, and the resulting ellipse-fit in blue.}
\label{lasco_c2_fig}
\end{figure}

\begin{figure}[!t]
\centering{\includegraphics[clip=true, scale=0.65, trim=0 0 0 0]{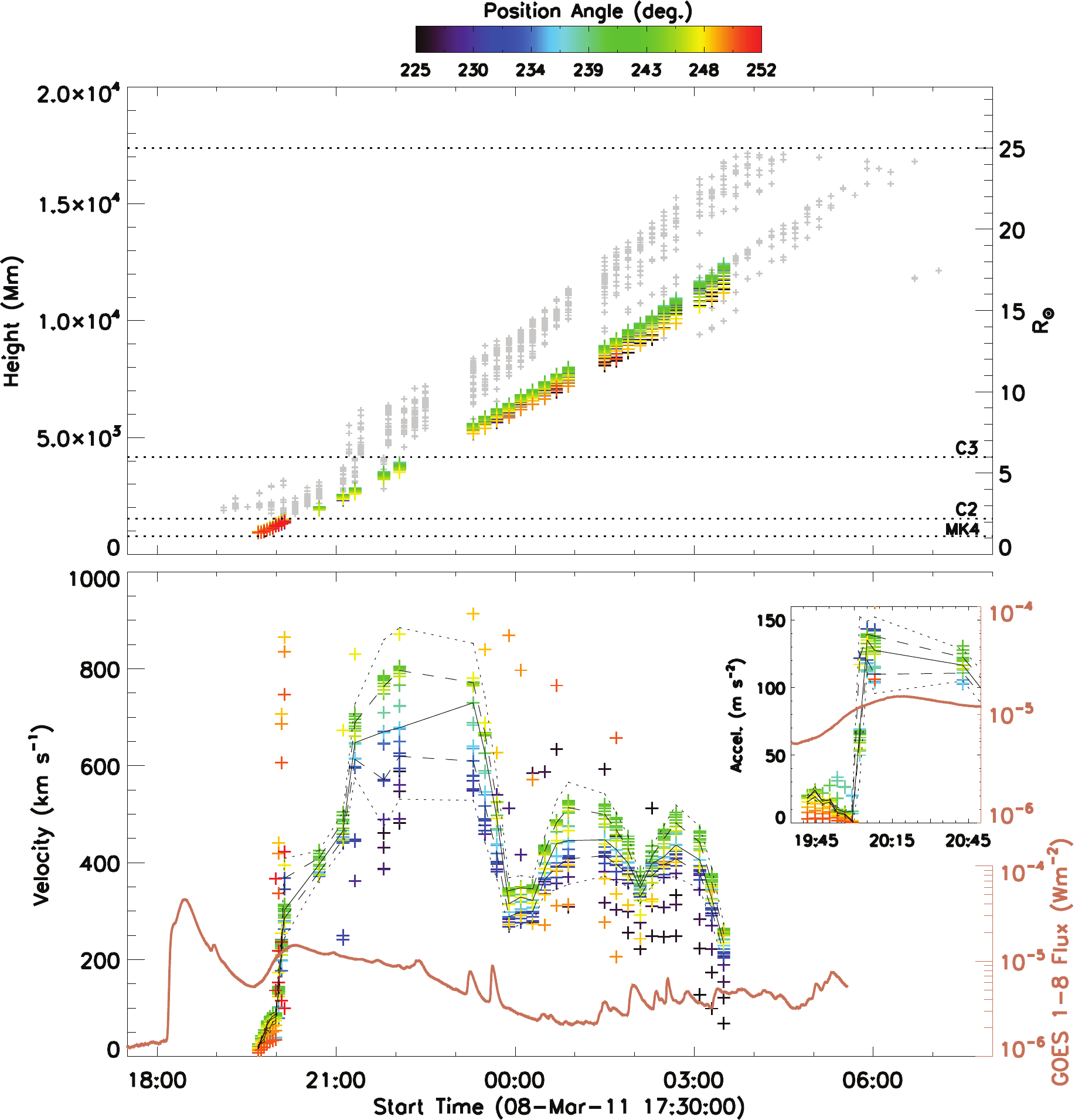}}
\caption{The kinematic profiles of height, velocity, and (inset) the early acceleration phase of the CME core material, detected and characterized via the multiscale edge enhancement and ellipse-fits (see Section\,\ref{sect:techniques} and Figs.\,\ref{ell_heights_inner} and \ref{lasco_c2_fig}). The automated CORIMP CME detections provide the height-time measurements of the CME front shown here in gray for reference, with the height-time measurements of the CME core  plotted in color according to their position angle. The fields-of-view of the Mk4, C2 and C3 instruments are indicated by the horizontal dotted lines, covering a useable range of 1.1\,--\,25\,$R_{\odot}$, with the CME core tracked to approximately 18\,$R_{\odot}$. A Savitzky-Golay filter was applied to the height-time measurements to obtain distribution profiles of velocity and acceleration, with the median, interquartiles range, and upper/lower fences over-plotted in solid, dashed and dotted lines respectively. Overlaid is the GOES X-ray 1\,--\,8\,{\AA} flux profile, showing the double-eruption peaks at about 18:28 and 20:15\,UT, the latter of which coincides with the CME jerk (abrupt increase in acceleration; see inset).}
\label{kins_CMEcore}
\end{figure}

The CME was observed to have a typical three-part structure that propagated out through the corona with a bulk speed in the range $\sim$\,400\,--\,600$\,km\,s^{-1}$ (based on the kinematics of the CME front detected and tracked in the CORIMP catalog\footnote{http://alshamess.ifa.hawaii.edu/CORIMP/}). The core of the CME was manually tracked via the multiscale methods and ellipse-fits discussed above (an example LASCO image is shown in Fig.\,\ref{lasco_c2_fig}). The resulting kinematics are plotted in Fig.\,\ref{kins_CMEcore}, where the height-time measurements of the CME core, observed with Mk4, C2 and C3, are shown in color (corresponding to the position angle of the measurements across the plane-of-sky) overlaid on the CME front height-time measurements shown in gray for reference. The Savitzky-Golay filter is used to derive the velocity and acceleration profiles for the CME core; whereby a distribution of velocity and acceleration values is obtained at each data point, with the corresponding median, interquartile range, and upper and lower fences overlaid on each profile as solid, dashed and dotted lines respectively (see \opencite{2013A&A...557A..96B} for a details). The inset acceleration phase of the CME core shows initial values of approximately $\sim$\,20$\,m\,s^{-2}$ increasing to $\sim$\,130$\,m\,s^{-2}$, referred to as the CME jerk by \inlinecite{2008ApJ...674..586S}. The steepness of this increase may in part be attributed to the numerical effects of the data gap between Mk4 and C2, where the spline-fit of the Savitzky-Golay filter (operating on a moving window of 6 neighboring data points) compensates somewhat for the increase in velocity that occurs between these two fields-of-view. The magnitude of the velocity, however, was verified by inspection of the different profiles within the separate instrument fields-of-view, for different degrees of polynomial fits and Savitzky-Golay filter sizes. That is, the rise in velocity was determined to be real, though its true steepness might be better quantified if it were possible to obtain a more complete set of measurements. Nevertheless, it occurs in sync with the second rise in the X-ray flare profile, being an indication of a very fast energy release that allowed the explosive eruption of the CME to begin, before later attaining a speed akin to the local solar wind speed. This double acceleration profile is further evidence for the complexity of the system whose morphology has been observed to change dramatically over the course of its evolution \cite{2012ApJ...746L...5S,2013ApJS..206...19M}.

\section{Discussion}
\label{sect:discussion}

The study of this particular event is especially interesting in the context of the flare-CME relationship. The two-stage flaring profile of the erupting loop system is evidence for a secondary heating process \cite{2012ApJ...746L...5S}, indicating two stages of magnetic reconnection that occur to first change the topology of the system and then allow for the subsequent flux rope eruption. This scenario demonstrates that a loss of stability occurred initially, to allow the loop system to rise and alter its magnetic configuration with an explosive energy release detected as an M-class flare. This was followed by a secondary energy release that allowed the underlying flux rope to erupt through the corona as a CME, with the production of a second X-ray peak and post-flare loops at the CME footpoints. In a similar study, \inlinecite{2006ApJ...651.1238Z} combine EIT and LASCO observations of the initiation phase of two successive prominence-eruption CMEs, and determine that they are driven by the kink instability and mass drainage with an impulsive acceleration onset resulting from magnetic reconnection beneath the filament.

It is not wholly clear why we observe different rates of motion of the material in the SWAP and Mk4 images during the CME onset (Fig.~\ref{ell_heights_inner}). One reason is possibly that the material observed by the different instruments corresponds to different parts of the erupting structure at different temperatures, which are difficult to dissociate from each other on the plane-of-sky. \inlinecite{2012ApJ...750...44B} observed a very similar effect for the four AIA passbands they used to track an erupting plasmoid, and demonstrated that different temperature structures do exhibit different eruption speeds (see, in particular, Fig.\,4 of their paper). The effect may further be attributed to the greater loss of signal in the EUV than the white-light as the eruption proceeds to the edge of the field-of-view (diminishing in its signal-to-noise ratio while the white-light may also be showing plasma pile-up); though this is difficult to reconcile with the observations that appear to show consistent loop structures as characterized in our analysis. Indeed if the offset is true and the different parts of the structure did undergo such differing rates of propagation on the same plane-of-sky, it may be that the trailing part simply became a different portion of the main CME and/or underwent a delayed jerk in its motion, which is not observed as the EUV signal diminishes towards the edge of the SWAP field-of-view. This would imply that there was some form of delayed or staggered eruption occurring throughout the CME structure, or that an expansion effect took over as the CME bubble formed, that created the observed offset between the different observations.


Either way, this event is intriguing in its possible interpretation. The post-flare loops in the first stage may be likened to the typical ones in the standard flare model (CSHKP: \opencite{1964NASSP..50..451C}; \opencite{1966Natur.211..695S}; \opencite{1974SoPh...34..323H}; \opencite{1976SoPh...50...85K}). Thus the early phase must be driven, at least in part, by a non-ideal process. The erupting loops, which were oriented north-south (above east-west flare ribbons seen in EUVI-Ahead), were then impeded by overlying field lines as seen in AIA data, causing a $\sim$\,50~minute stall in the eruption. For the eruption to proceed, the upward magnetic pressure force must overcome the downward magnetic tension force, following either an energy build-up and/or a topological change to the system during the continued reconnection in the decay phase of the first stage. The jerk in the CME motion when it quickly accelerates during the second stage flare is evidence of this catastrophic energy release.

The exact physical connection between these two stages is unclear. For example, one possibility we investigated is that of tension reduction, through a mechanism such as magnetic breakout whereby the overlying arcades are removed by coronal reconnection. This was motivated by two factors. Firstly, the SDO/HMI magnetogram data in the days preceding the eruption show the active region evolving into a possible multi-polar configuration. But since the magnetogram data on the limb is not clear (i.e., we cannot directly observe or extrapolate the active region topology) this does not provide strong evidence. Secondly, \inlinecite{2009ApJ...693.1178V} demonstrate the ``bugle effect" in a breakout model that pertains to a streamer blowout CME like this one. This follows the observations of \inlinecite{2013ApJS..206...19M} that show a streamer swelling on the limb above the active region to large heights in the LASCO coronagraphs, which evolved slowly over the preceding days in the lead-up to the eruptive event. Helmet streamers like this are sometimes observed prior to CME initiation \cite{1993JGR....9813177H} though not always to such great heights due to the low signal-to-noise ratio -- an issue overcome by the dynamic separation technique of \inlinecite{2012ApJ...752..144M}. \inlinecite{2009ApJ...693.1178V} also describe how the rising flux rope can ``snow plow" the plasma ahead of it, with the effect of broadening the helmet streamer and facilitating reconnection between the flux rope and the helmet field. Flare reconnection then sets in due to the expansion of the magnetic field in the wake of the CME eruption as indicated by the elongated ``X"-shaped structure and post-flare loops in the observations of \inlinecite{2012ApJ...746L...5S}.

However, since the presence of a quadrupolar topology required for breakout reconnection is not clear, we considered the formation of a kink or torus-unstable flux rope \cite{2004A&A...413L..27T,2010ApJ...708..314A}, with evidence of internal tether-cutting during the eruption (c.f., \opencite{2010ApJ...721.1579R}). \inlinecite{2014zuccarello} observed evidence of line-tying reconnection and the restructuring of a filament that facilitated its eruption due to the torus-instability. Since here we observe a rather unique ``two-stage" event, it may be posited that a facilitator (such as tension reduction via tether-cutting reconnection, arcade shearing, flux emergence or cancellation) allowed for a loss-of-equilibrium in the first stage, and a subsequent instability then triggered a catastrophic eruption in the second stage. In this case, the overlying magnetic field tension was strong enough, especially in the presence of the helmet streamer, to stall the eruption until the upward magnetic pressure rose to a critical point. \inlinecite{2004A&A...413L..27T} studied the stability of the loop model by \inlinecite{1999A&A...351..707T} in a twisted configuration that leads to an ideal kink instability, where the resulting vertical current sheet corresponds to the central element of the standard flare model. They note that the surrounding potential field would prevent a global eruption unless the overlying field is sufficiently weakened or magnetic reconnection occurs in either the formed current sheet or with neighboring flux systems (requiring a multiple-loop configuration). For this event, although the X-ray flare profile indicates a non-ideal process from the start, we may be seeing a similar form of halted eruption before a second stage process kicks in that is dominated by a kink-instability.

\inlinecite{2010ApJ...708..314A} consider the non-ideal case via a flux cancellation model that transforms a sheared arcade into a slowly rising and stable flux rope. Eventually an altitude is reached at which the rate of overlying field decay exceeds the upward magnetic pressure and a torus instability causes a rapid eruption in the form of a CME (of inverse tear-drop shape as observed in this event). Furthermore, they describe how part of the flux rope in the early stages rises faster, due to an asymmetric build-up of the pre-eruptive flux rope. Then the system enters a phase of fast expansion (accelerating to a velocity of $400\,km\,s^{-1}$ in their model), where the overlying field lines that are not affected by the shearing motions correspond to the front of the CME commonly tracked in coronagraph observations. These modeled effects may offer an explanation for the observations of the 8~March~2011 event here, although without any measure of the decay index of the background magnetic field we cannot be certain that a torus-instability is what we definitively observe.

\section{Conclusions}
\label{sect:conclusions}

The \emph{PROBA2}/SWAP imager is unique in that it provides radially extended EUV observations of the Sun and low-corona to greater heights than other EUV imagers such as \emph{SDO}/AIA. An ongoing goal in solar physics has been to study the connection between processes on the Sun and the effects felt elsewhere in the heliosphere; a connection known to lie predominantly between the regions of the photosphere, chromosphere and corona. Therefore, obtaining extensive observations across the solar atmosphere is paramount to understanding the physics at play. Since the LASCO/C1 coronagraph was lost very early on in the \emph{SOHO} mission, observations of the low corona have generally been quite limited. In order to bridge this gap and garner some knowledge of the low-coronal initiation phase of CMEs, we have combined the SWAP observations with those of the ground-based Mk4 coronameter, to directly compare the EUV and white-light imagery. This was achieved through the use of advanced image processing techniques to overcome the low signal-to-noise ratio in these data and characterize the erupting structures of interest. The subsequent investigation of the dynamics of a specific case-study on 8~March~2011 provides insight to the early CME formation and eruption, using SWAP and Mk4 in tandem with the observations of AIA and LASCO. This compliments the previous investigation of \inlinecite{2012ApJ...746L...5S} who first reported on the event's two-stage flaring profile as evidence for secondary heating. It is concluded that this event shows evidence for either a kink-unstable or torus-unstable flux rope that is first formed by possible shearing motions and flux cancellation, before undergoing a magnetic pressure build-up and overlying tension decay that drives a subsequent fast CME and resulting streamer blowout. 

Our study highlights the importance of multi-wavelength, high-cadence, extensive coverage observations of the low-corona, where the physical mechanisms that underly CME formation, initiation, and connection with flares and surrounding coronal field may best be investigated. It is hoped that future instruments such as the new COSMO K-coronagraph\footnote{http://www.cosmo.ucar.edu/kcoronagraph.html} will help to further advance our knowledge of these mechanisms.


%

%

%

%
 \begin{acks}
 
This work is supported by SHINE grant 0962716 and NASA grants NNX08AJ07G and NNX13AG11G to the Institute for Astronomy.
SWAP is a project of the Centre Spatial de Li'ege and the Royal Observatory of Belgium funded by the Belgian Federal Science Policy Office (BELSPO).
Mk4 data is provided courtesy of the Mauna Loa Solar Observatory, operated by the High Altitude Observatory, as part of the National Center for Atmospheric Research (NCAR). NCAR is supported by the National Science Foundation.
The \emph{SOHO}/LASCO data used here are produced by a consortium of the Naval Research Laboratory (USA), Max-Planck-Institut fuer Aeronomie (Germany), Laboratoire d'Astronomie (France), and the University of Birmingham (UK). SOHO is a project of international cooperation between ESA and NASA.
\emph{SDO} data supplied courtesy of the NASA/\emph{SDO} consortia. The authors thank the anonymous referee for their helpful comments. JPB is grateful to have been a PROBA2 Guest Investigator.

 \end{acks}

%
%
 \bibliographystyle{spr-mp-sola.bst}
 \bibliography{references.bib}  

\begin{thebibliography}{70}
\ifx \bisbn   \undefined \def \bisbn  #1{ISBN #1}\fi
\ifx \binits  \undefined \def \binits#1{#1}\fi
\ifx \bauthor  \undefined \def \bauthor#1{#1}\fi
\ifx \batitle  \undefined \def \batitle#1{#1}\fi
\ifx \bjtitle  \undefined \def \bjtitle#1{\textit{#1}}\fi
\ifx \bvolume  \undefined \def \bvolume#1{\textbf{#1}}\fi
\ifx \byear  \undefined \def \byear#1{#1}\fi
\ifx \bissue  \undefined \def \bissue#1{#1}\fi
\ifx \bfpage  \undefined \def \bfpage#1{#1}\fi
\ifx \blpage  \undefined \def \blpage #1{#1}\fi
\ifx \burl  \undefined \def \burl#1{\textsf{#1}}\fi
\ifx \href  \undefined \def \href#1#2{\textsf{#2}}\fi
\ifx \doiurl  \undefined \def
  \doiurl#1{\href{http://dx.doi.org/#1}{\textsf{#1}}}\fi
\ifx \betal  \undefined \def \betal{\textit{et al.}}\fi
\ifx \binstitute  \undefined \def \binstitute#1{#1}\fi
\ifx \bctitle  \undefined \def \bctitle#1{#1}\fi
\ifx \beditor  \undefined \def \beditor#1{#1}\fi
\ifx \bpublisher  \undefined \def \bpublisher#1{#1}\fi
\ifx \bbtitle  \undefined \def \bbtitle#1{\textit{#1}}\fi
\ifx \bedition  \undefined \def \bedition#1{#1}\fi
\ifx \bseriesno  \undefined \def \bseriesno#1{\textbf{#1}}\fi
\ifx \blocation  \undefined \def \blocation#1{#1}\fi
\ifx \bsertitle  \undefined \def \bsertitle#1{\textit{#1}}\fi
\ifx \bsnm \undefined \def \bsnm#1{#1}\fi
\ifx \bsuffix \undefined \def \bsuffix#1{#1}\fi
\ifx \bparticle \undefined \def \bparticle#1{#1}\fi
\ifx \barticle \undefined \def \barticle#1{}\fi
\ifx \botherref \undefined \def \botherref#1{}\fi
\ifx \url \undefined \def \url#1{\textsf{#1}}\fi
\ifx \bchapter \undefined \def \bchapter#1{}\fi
\ifx \bbook \undefined \def \bbook#1{}\fi
\ifx \bcomment \undefined \def \bcomment#1{#1}\fi
\ifx \oauthor \undefined \def \oauthor#1{#1}\fi
\ifx \citeauthoryear \undefined \def \citeauthoryear#1{#1}\fi
\def \endbibitem {}
\ifx \bconflocation  \undefined \def \bconflocation#1{#1} \fi

\bibitem[\protect\citeauthoryear{{Amari}
  \textit{et~al.}}{2003}]{2003ApJ...595.1231A}
\begin{barticle}
\bauthor{\bsnm{{Amari}}, \binits{T.}},
\bauthor{\bsnm{{Luciani}}, \binits{J.F.}},
\bauthor{\bsnm{{Aly}}, \binits{J.J.}},
\bauthor{\bsnm{{Mikic}}, \binits{Z.}},
\bauthor{\bsnm{{Linker}}, \binits{J.}}:
\byear{2003},
\batitle{{Coronal Mass Ejection: Initiation, Magnetic Helicity, and Flux Ropes.
  II. Turbulent Diffusion-driven Evolution}}.
\bjtitle{\apj}
\bvolume{595},
\bfpage{1231}\,--\,\blpage{1250}.
doi:\doiurl{10.1086/377444}.
\end{barticle}
\endbibitem

\bibitem[\protect\citeauthoryear{{Antiochos}, {DeVore}, and
  {Klimchuk}}{1999}]{1999ApJ...510..485A}
\begin{barticle}
\bauthor{\bsnm{{Antiochos}}, \binits{S.K.}},
\bauthor{\bsnm{{DeVore}}, \binits{C.R.}},
\bauthor{\bsnm{{Klimchuk}}, \binits{J.A.}}:
\byear{1999},
\batitle{{A Model for Solar Coronal Mass Ejections}}.
\bjtitle{\apj}
\bvolume{510},
\bfpage{485}\,--\,\blpage{493}.
doi:\doiurl{10.1086/306563}.
\end{barticle}
\endbibitem

\bibitem[\protect\citeauthoryear{{Aulanier}
  \textit{et~al.}}{2010}]{2010ApJ...708..314A}
\begin{barticle}
\bauthor{\bsnm{{Aulanier}}, \binits{G.}},
\bauthor{\bsnm{{T{\"o}r{\"o}k}}, \binits{T.}},
\bauthor{\bsnm{{D{\'e}moulin}}, \binits{P.}},
\bauthor{\bsnm{{DeLuca}}, \binits{E.E.}}:
\byear{2010},
\batitle{{Formation of Torus-Unstable Flux Ropes and Electric Currents in
  Erupting Sigmoids}}.
\bjtitle{\apj}
\bvolume{708},
\bfpage{314}\,--\,\blpage{333}.
doi:\doiurl{10.1088/0004-637X/708/1/314}.
\end{barticle}
\endbibitem

\bibitem[\protect\citeauthoryear{{Bain}
  \textit{et~al.}}{2012}]{2012ApJ...750...44B}
\begin{barticle}
\bauthor{\bsnm{{Bain}}, \binits{H.M.}},
\bauthor{\bsnm{{Krucker}}, \binits{S.}},
\bauthor{\bsnm{{Glesener}}, \binits{L.}},
\bauthor{\bsnm{{Lin}}, \binits{R.P.}}:
\byear{2012},
\batitle{{Radio Imaging of Shock-accelerated Electrons Associated with an
  Erupting Plasmoid on 2010 November 3}}.
\bjtitle{\apj}
\bvolume{750},
\bfpage{44}.
doi:\doiurl{10.1088/0004-637X/750/1/44}.
\end{barticle}
\endbibitem

\bibitem[\protect\citeauthoryear{{Brueckner}
  \textit{et~al.}}{1995}]{1995SoPh..162..357B}
\begin{barticle}
\bauthor{\bsnm{{Brueckner}}, \binits{G.E.}},
\bauthor{\bsnm{{Howard}}, \binits{R.A.}},
\bauthor{\bsnm{{Koomen}}, \binits{M.J.}},
\bauthor{\bsnm{{Korendyke}}, \binits{C.M.}},
\bauthor{\bsnm{{Michels}}, \binits{D.J.}},
\bauthor{\bsnm{{Moses}}, \binits{J.D.}},
\bauthor{\bsnm{{Socker}}, \binits{D.G.}},
\bauthor{\bsnm{{Dere}}, \binits{K.P.}},
\bauthor{\bsnm{{Lamy}}, \binits{P.L.}},
\bauthor{\bsnm{{Llebaria}}, \binits{A.}},
\bauthor{\bsnm{{Bout}}, \binits{M.V.}},
\bauthor{\bsnm{{Schwenn}}, \binits{R.}},
\bauthor{\bsnm{{Simnett}}, \binits{G.M.}},
\bauthor{\bsnm{{Bedford}}, \binits{D.K.}},
\bauthor{\bsnm{{Eyles}}, \binits{C.J.}}:
\byear{1995},
\batitle{{The Large Angle Spectroscopic Coronagraph (LASCO)}}.
\bjtitle{\solphys}
\bvolume{162},
\bfpage{357}\,--\,\blpage{402}.
doi:\doiurl{10.1007/BF00733434}.
\end{barticle}
\endbibitem

\bibitem[\protect\citeauthoryear{{Byrne}
  \textit{et~al.}}{2009}]{2009A&A...495..325B}
\begin{barticle}
\bauthor{\bsnm{{Byrne}}, \binits{J.P.}},
\bauthor{\bsnm{{Gallagher}}, \binits{P.T.}},
\bauthor{\bsnm{{McAteer}}, \binits{R.T.J.}},
\bauthor{\bsnm{{Young}}, \binits{C.A.}}:
\byear{2009},
\batitle{{The kinematics of coronal mass ejections using multiscale methods}}.
\bjtitle{\aap}
\bvolume{495},
\bfpage{325}\,--\,\blpage{334}.
doi:\doiurl{10.1051/0004-6361:200809811}.
\end{barticle}
\endbibitem

\bibitem[\protect\citeauthoryear{{Byrne}
  \textit{et~al.}}{2010}]{2010NatCo...1E..74B}
\begin{barticle}
\bauthor{\bsnm{{Byrne}}, \binits{J.P.}},
\bauthor{\bsnm{{Maloney}}, \binits{S.A.}},
\bauthor{\bsnm{{McAteer}}, \binits{R.T.J.}},
\bauthor{\bsnm{{Refojo}}, \binits{J.M.}},
\bauthor{\bsnm{{Gallagher}}, \binits{P.T.}}:
\byear{2010},
\batitle{{Propagation of an Earth-directed coronal mass ejection in three
  dimensions}}.
\bjtitle{Nature Communications}
\bvolume{1}.
doi:\doiurl{10.1038/ncomms1077}.
\end{barticle}
\endbibitem

\bibitem[\protect\citeauthoryear{{Byrne}
  \textit{et~al.}}{2012}]{2012ApJ...752..145B}
\begin{barticle}
\bauthor{\bsnm{{Byrne}}, \binits{J.P.}},
\bauthor{\bsnm{{Morgan}}, \binits{H.}},
\bauthor{\bsnm{{Habbal}}, \binits{S.R.}},
\bauthor{\bsnm{{Gallagher}}, \binits{P.T.}}:
\byear{2012},
\batitle{{Automatic Detection and Tracking of Coronal Mass Ejections. II.
  Multiscale Filtering of Coronagraph Images}}.
\bjtitle{\apj}
\bvolume{752},
\bfpage{145}.
doi:\doiurl{10.1088/0004-637X/752/2/145}.
\end{barticle}
\endbibitem

\bibitem[\protect\citeauthoryear{{Byrne}
  \textit{et~al.}}{2013}]{2013A&A...557A..96B}
\begin{barticle}
\bauthor{\bsnm{{Byrne}}, \binits{J.P.}},
\bauthor{\bsnm{{Long}}, \binits{D.M.}},
\bauthor{\bsnm{{Gallagher}}, \binits{P.T.}},
\bauthor{\bsnm{{Bloomfield}}, \binits{D.S.}},
\bauthor{\bsnm{{Maloney}}, \binits{S.A.}},
\bauthor{\bsnm{{McAteer}}, \binits{R.T.J.}},
\bauthor{\bsnm{{Morgan}}, \binits{H.}},
\bauthor{\bsnm{{Habbal}}, \binits{S.R.}}:
\byear{2013},
\batitle{{Improved methods for determining the kinematics of coronal mass
  ejections and coronal waves}}.
\bjtitle{\aap}
\bvolume{557},
\bfpage{A96}.
doi:\doiurl{10.1051/0004-6361/201321223}.
\end{barticle}
\endbibitem

\bibitem[\protect\citeauthoryear{{Carley}
  \textit{et~al.}}{2013}]{2013NatPh...9..811C}
\begin{barticle}
\bauthor{\bsnm{{Carley}}, \binits{E.P.}},
\bauthor{\bsnm{{Long}}, \binits{D.M.}},
\bauthor{\bsnm{{Byrne}}, \binits{J.P.}},
\bauthor{\bsnm{{Zucca}}, \binits{P.}},
\bauthor{\bsnm{{Bloomfield}}, \binits{D.S.}},
\bauthor{\bsnm{{McCauley}}, \binits{J.}},
\bauthor{\bsnm{{Gallagher}}, \binits{P.T.}}:
\byear{2013},
\batitle{{Quasiperiodic acceleration of electrons by a plasmoid-driven shock in
  the solar atmosphere}}.
\bjtitle{Nature Physics}
\bvolume{9},
\bfpage{811}\,--\,\blpage{816}.
doi:\doiurl{10.1038/nphys2767}.
\end{barticle}
\endbibitem

\bibitem[\protect\citeauthoryear{{Carmichael}}{1964}]{1964NASSP..50..451C}
\begin{barticle}
\bauthor{\bsnm{{Carmichael}}, \binits{H.}}:
\byear{1964},
\batitle{{A Process for Flares}}.
\bjtitle{NASA Special Publication}
\bvolume{50},
\bfpage{451}.
\end{barticle}
\endbibitem

\bibitem[\protect\citeauthoryear{{Chen}}{1996}]{1996JGR...10127499C}
\begin{barticle}
\bauthor{\bsnm{{Chen}}, \binits{J.}}:
\byear{1996},
\batitle{{Theory of prominence eruption and propagation: Interplanetary
  consequences}}.
\bjtitle{\jgr}
\bvolume{101},
\bfpage{27499}\,--\,\blpage{27520}.
doi:\doiurl{10.1029/96JA02644}.
\end{barticle}
\endbibitem

\bibitem[\protect\citeauthoryear{{Chen}}{2011}]{2011LRSP....8....1C}
\begin{barticle}
\bauthor{\bsnm{{Chen}}, \binits{P.F.}}:
\byear{2011},
\batitle{{Coronal Mass Ejections: Models and Their Observational Basis}}.
\bjtitle{Living Reviews in Solar Physics}
\bvolume{8},
\bfpage{1}.
\end{barticle}
\endbibitem

\bibitem[\protect\citeauthoryear{{Cremades} and
  {Bothmer}}{2004}]{2004A&A...422..307C}
\begin{barticle}
\bauthor{\bsnm{{Cremades}}, \binits{H.}},
\bauthor{\bsnm{{Bothmer}}, \binits{V.}}:
\byear{2004},
\batitle{{On the three-dimensional configuration of coronal mass ejections}}.
\bjtitle{\aap}
\bvolume{422},
\bfpage{307}\,--\,\blpage{322}.
doi:\doiurl{10.1051/0004-6361:20035776}.
\end{barticle}
\endbibitem

\bibitem[\protect\citeauthoryear{{Dauphin}, {Vilmer}, and
  {Krucker}}{2006}]{2006A&A...455..339D}
\begin{barticle}
\bauthor{\bsnm{{Dauphin}}, \binits{C.}},
\bauthor{\bsnm{{Vilmer}}, \binits{N.}},
\bauthor{\bsnm{{Krucker}}, \binits{S.}}:
\byear{2006},
\batitle{{Observations of a soft X-ray rising loop associated with a type II
  burst and a coronal mass ejection in the 03 November 2003 X-ray flare}}.
\bjtitle{\aap}
\bvolume{455},
\bfpage{339}\,--\,\blpage{348}.
doi:\doiurl{10.1051/0004-6361:20054535}.
\end{barticle}
\endbibitem

\bibitem[\protect\citeauthoryear{{Domingo}, {Fleck}, and
  {Poland}}{1995}]{1995SoPh..162....1D}
\begin{barticle}
\bauthor{\bsnm{{Domingo}}, \binits{V.}},
\bauthor{\bsnm{{Fleck}}, \binits{B.}},
\bauthor{\bsnm{{Poland}}, \binits{A.I.}}:
\byear{1995},
\batitle{{The SOHO Mission: an Overview}}.
\bjtitle{\solphys}
\bvolume{162},
\bfpage{1}\,--\,\blpage{2}.
doi:\doiurl{10.1007/BF00733425}.
\end{barticle}
\endbibitem

\bibitem[\protect\citeauthoryear{{Druckm{\"u}llerov{\'a}}, {Morgan}, and
  {Habbal}}{2011}]{2011ApJ...737...88D}
\begin{barticle}
\bauthor{\bsnm{{Druckm{\"u}llerov{\'a}}}, \binits{H.}},
\bauthor{\bsnm{{Morgan}}, \binits{H.}},
\bauthor{\bsnm{{Habbal}}, \binits{S.R.}}:
\byear{2011},
\batitle{{Enhancing Coronal Structures with the Fourier
  Normalizing-radial-graded Filter}}.
\bjtitle{\apj}
\bvolume{737},
\bfpage{88}.
doi:\doiurl{10.1088/0004-637X/737/2/88}.
\end{barticle}
\endbibitem

\bibitem[\protect\citeauthoryear{{Elmore}
  \textit{et~al.}}{2003}]{2003SPIE.4843...66E}
\begin{bchapter}
\bauthor{\bsnm{{Elmore}}, \binits{D.F.}},
\bauthor{\bsnm{{Burkepile}}, \binits{J.T.}},
\bauthor{\bsnm{{Darnell}}, \binits{J.A.}},
\bauthor{\bsnm{{Lecinski}}, \binits{A.R.}},
\bauthor{\bsnm{{Stanger}}, \binits{A.L.}}:
\byear{2003},
\bctitle{{Calibration of a ground-based solar coronal polarimeter}}.
In: \beditor{\bsnm{{Fineschi}}, \binits{S.}} (ed.)
\bbtitle{Society of Photo-Optical Instrumentation Engineers (SPIE) Conference
  Series},
\bsertitle{Society of Photo-Optical Instrumentation Engineers (SPIE) Conference
  Series}
\bseriesno{4843},
\bfpage{66}\,--\,\blpage{75}.
doi:\doiurl{10.1117/12.459279}.
\end{bchapter}
\endbibitem

\bibitem[\protect\citeauthoryear{{Filippov} and
  {Koutchmy}}{2008}]{2008AnGeo..26.3025F}
\begin{barticle}
\bauthor{\bsnm{{Filippov}}, \binits{B.}},
\bauthor{\bsnm{{Koutchmy}}, \binits{S.}}:
\byear{2008},
\batitle{{Causal relationships between eruptive prominences and coronal mass
  ejections}}.
\bjtitle{Annales Geophysicae}
\bvolume{26},
\bfpage{3025}\,--\,\blpage{3031}.
doi:\doiurl{10.5194/angeo-26-3025-2008}.
\end{barticle}
\endbibitem

\bibitem[\protect\citeauthoryear{{Forbes} and
  {Priest}}{1995}]{1995ApJ...446..377F}
\begin{barticle}
\bauthor{\bsnm{{Forbes}}, \binits{T.G.}},
\bauthor{\bsnm{{Priest}}, \binits{E.R.}}:
\byear{1995},
\batitle{{Photospheric Magnetic Field Evolution and Eruptive Flares}}.
\bjtitle{\apj}
\bvolume{446},
\bfpage{377}.
doi:\doiurl{10.1086/175797}.
\end{barticle}
\endbibitem

\bibitem[\protect\citeauthoryear{{Gallagher}
  \textit{et~al.}}{2011}]{2011AdSpR..47.2118G}
\begin{barticle}
\bauthor{\bsnm{{Gallagher}}, \binits{P.T.}},
\bauthor{\bsnm{{Young}}, \binits{C.A.}},
\bauthor{\bsnm{{Byrne}}, \binits{J.P.}},
\bauthor{\bsnm{{McAteer}}, \binits{R.T.J.}}:
\byear{2011},
\batitle{{Coronal mass ejection detection using wavelets, curvelets and
  ridgelets: Applications for space weather monitoring}}.
\bjtitle{Advances in Space Research}
\bvolume{47},
\bfpage{2118}\,--\,\blpage{2126}.
doi:\doiurl{10.1016/j.asr.2010.03.028}.
\end{barticle}
\endbibitem

\bibitem[\protect\citeauthoryear{{Gibson}
  \textit{et~al.}}{2006}]{2006ApJ...641..590G}
\begin{barticle}
\bauthor{\bsnm{{Gibson}}, \binits{S.E.}},
\bauthor{\bsnm{{Foster}}, \binits{D.}},
\bauthor{\bsnm{{Burkepile}}, \binits{J.}},
\bauthor{\bsnm{{de Toma}}, \binits{G.}},
\bauthor{\bsnm{{Stanger}}, \binits{A.}}:
\byear{2006},
\batitle{{The Calm before the Storm: The Link between Quiescent Cavities and
  Coronal Mass Ejections}}.
\bjtitle{\apj}
\bvolume{641},
\bfpage{590}\,--\,\blpage{605}.
doi:\doiurl{10.1086/500446}.
\end{barticle}
\endbibitem

\bibitem[\protect\citeauthoryear{{Gopalswamy}
  \textit{et~al.}}{2003}]{2003ApJ...586..562G}
\begin{barticle}
\bauthor{\bsnm{{Gopalswamy}}, \binits{N.}},
\bauthor{\bsnm{{Shimojo}}, \binits{M.}},
\bauthor{\bsnm{{Lu}}, \binits{W.}},
\bauthor{\bsnm{{Yashiro}}, \binits{S.}},
\bauthor{\bsnm{{Shibasaki}}, \binits{K.}},
\bauthor{\bsnm{{Howard}}, \binits{R.A.}}:
\byear{2003},
\batitle{{Prominence Eruptions and Coronal Mass Ejection: A Statistical Study
  Using Microwave Observations}}.
\bjtitle{\apj}
\bvolume{586},
\bfpage{562}\,--\,\blpage{578}.
doi:\doiurl{10.1086/367614}.
\end{barticle}
\endbibitem

\bibitem[\protect\citeauthoryear{{Halain}
  \textit{et~al.}}{2013}]{2013SoPh..286...67H}
\begin{barticle}
\bauthor{\bsnm{{Halain}}, \binits{J.-P.}},
\bauthor{\bsnm{{Berghmans}}, \binits{D.}},
\bauthor{\bsnm{{Seaton}}, \binits{D.B.}},
\bauthor{\bsnm{{Nicula}}, \binits{B.}},
\bauthor{\bsnm{{De Groof}}, \binits{A.}},
\bauthor{\bsnm{{Mierla}}, \binits{M.}},
\bauthor{\bsnm{{Mazzoli}}, \binits{A.}},
\bauthor{\bsnm{{Defise}}, \binits{J.-M.}},
\bauthor{\bsnm{{Rochus}}, \binits{P.}}:
\byear{2013},
\batitle{{The SWAP EUV Imaging Telescope. Part II: In-flight Performance and
  Calibration}}.
\bjtitle{\solphys}
\bvolume{286},
\bfpage{67}\,--\,\blpage{91}.
doi:\doiurl{10.1007/s11207-012-0183-6}.
\end{barticle}
\endbibitem

\bibitem[\protect\citeauthoryear{{Hirayama}}{1974}]{1974SoPh...34..323H}
\begin{barticle}
\bauthor{\bsnm{{Hirayama}}, \binits{T.}}:
\byear{1974},
\batitle{{Theoretical Model of Flares and Prominences. I: Evaporating Flare
  Model}}.
\bjtitle{\solphys}
\bvolume{34},
\bfpage{323}\,--\,\blpage{338}.
doi:\doiurl{10.1007/BF00153671}.
\end{barticle}
\endbibitem

\bibitem[\protect\citeauthoryear{{Howard}
  \textit{et~al.}}{2008}]{2008SSRv..136...67H}
\begin{barticle}
\bauthor{\bsnm{{Howard}}, \binits{R.A.}},
\bauthor{\bsnm{{Moses}}, \binits{J.D.}},
\bauthor{\bsnm{{Vourlidas}}, \binits{A.}},
\bauthor{\bsnm{{Newmark}}, \binits{J.S.}},
\bauthor{\bsnm{{Socker}}, \binits{D.G.}},
\bauthor{\bsnm{{Plunkett}}, \binits{S.P.}},
\bauthor{\bsnm{{Korendyke}}, \binits{C.M.}},
\bauthor{\bsnm{{Cook}}, \binits{J.W.}},
\bauthor{\bsnm{{Hurley}}, \binits{A.}},
\bauthor{\bsnm{{Davila}}, \binits{J.M.}},
\bauthor{\bsnm{{Thompson}}, \binits{W.T.}},
\bauthor{\bsnm{{St Cyr}}, \binits{O.C.}},
\bauthor{\bsnm{{Mentzell}}, \binits{E.}},
\bauthor{\bsnm{{Mehalick}}, \binits{K.}},
\bauthor{\bsnm{{Lemen}}, \binits{J.R.}},
\bauthor{\bsnm{{Wuelser}}, \binits{J.P.}},
\bauthor{\bsnm{{Duncan}}, \binits{D.W.}},
\bauthor{\bsnm{{Tarbell}}, \binits{T.D.}},
\bauthor{\bsnm{{Wolfson}}, \binits{C.J.}},
\bauthor{\bsnm{{Moore}}, \binits{A.}},
\bauthor{\bsnm{{Harrison}}, \binits{R.A.}},
\bauthor{\bsnm{{Waltham}}, \binits{N.R.}},
\bauthor{\bsnm{{Lang}}, \binits{J.}},
\bauthor{\bsnm{{Davis}}, \binits{C.J.}},
\bauthor{\bsnm{{Eyles}}, \binits{C.J.}},
\bauthor{\bsnm{{Mapson-Menard}}, \binits{H.}},
\bauthor{\bsnm{{Simnett}}, \binits{G.M.}},
\bauthor{\bsnm{{Halain}}, \binits{J.P.}},
\bauthor{\bsnm{{Defise}}, \binits{J.M.}},
\bauthor{\bsnm{{Mazy}}, \binits{E.}},
\bauthor{\bsnm{{Rochus}}, \binits{P.}},
\bauthor{\bsnm{{Mercier}}, \binits{R.}},
\bauthor{\bsnm{{Ravet}}, \binits{M.F.}},
\bauthor{\bsnm{{Delmotte}}, \binits{F.}},
\bauthor{\bsnm{{Auchere}}, \binits{F.}},
\bauthor{\bsnm{{Delaboudiniere}}, \binits{J.P.}},
\bauthor{\bsnm{{Bothmer}}, \binits{V.}},
\bauthor{\bsnm{{Deutsch}}, \binits{W.}},
\bauthor{\bsnm{{Wang}}, \binits{D.}},
\bauthor{\bsnm{{Rich}}, \binits{N.}},
\bauthor{\bsnm{{Cooper}}, \binits{S.}},
\bauthor{\bsnm{{Stephens}}, \binits{V.}},
\bauthor{\bsnm{{Maahs}}, \binits{G.}},
\bauthor{\bsnm{{Baugh}}, \binits{R.}},
\bauthor{\bsnm{{McMullin}}, \binits{D.}},
\bauthor{\bsnm{{Carter}}, \binits{T.}}:
\byear{2008},
\batitle{{Sun Earth Connection Coronal and Heliospheric Investigation
  (SECCHI)}}.
\bjtitle{Space Science Reviews}
\bvolume{136},
\bfpage{67}\,--\,\blpage{115}.
doi:\doiurl{10.1007/s11214-008-9341-4}.
\end{barticle}
\endbibitem

\bibitem[\protect\citeauthoryear{{Howard} and
  {Harrison}}{2013}]{2013SoPh..285..269H}
\begin{barticle}
\bauthor{\bsnm{{Howard}}, \binits{T.A.}},
\bauthor{\bsnm{{Harrison}}, \binits{R.A.}}:
\byear{2013},
\batitle{{Stealth Coronal Mass Ejections: A Perspective}}.
\bjtitle{\solphys}
\bvolume{285},
\bfpage{269}\,--\,\blpage{280}.
doi:\doiurl{10.1007/s11207-012-0217-0}.
\end{barticle}
\endbibitem

\bibitem[\protect\citeauthoryear{{Hundhausen}}{1993}]{1993JGR....9813177H}
\begin{barticle}
\bauthor{\bsnm{{Hundhausen}}, \binits{A.J.}}:
\byear{1993},
\batitle{{Sizes and locations of coronal mass ejections - SMM observations from
  1980 and 1984-1989}}.
\bjtitle{\jgr}
\bvolume{98},
\bfpage{13177}.
doi:\doiurl{10.1029/93JA00157}.
\end{barticle}
\endbibitem

\bibitem[\protect\citeauthoryear{{Illing} and
  {Hundhausen}}{1986}]{1986JGR....9110951I}
\begin{barticle}
\bauthor{\bsnm{{Illing}}, \binits{R.M.E.}},
\bauthor{\bsnm{{Hundhausen}}, \binits{A.J.}}:
\byear{1986},
\batitle{{Disruption of a coronal streamer by an eruptive prominence and
  coronal mass ejection}}.
\bjtitle{\jgr}
\bvolume{91},
\bfpage{10951}\,--\,\blpage{10960}.
doi:\doiurl{10.1029/JA091iA10p10951}.
\end{barticle}
\endbibitem

\bibitem[\protect\citeauthoryear{{Kaiser}
  \textit{et~al.}}{2008}]{2008SSRv..136....5K}
\begin{barticle}
\bauthor{\bsnm{{Kaiser}}, \binits{M.L.}},
\bauthor{\bsnm{{Kucera}}, \binits{T.A.}},
\bauthor{\bsnm{{Davila}}, \binits{J.M.}},
\bauthor{\bsnm{{St.~Cyr}}, \binits{O.C.}},
\bauthor{\bsnm{{Guhathakurta}}, \binits{M.}},
\bauthor{\bsnm{{Christian}}, \binits{E.}}:
\byear{2008},
\batitle{{The STEREO Mission: An Introduction}}.
\bjtitle{Space Science Reviews}
\bvolume{136},
\bfpage{5}\,--\,\blpage{16}.
doi:\doiurl{10.1007/s11214-007-9277-0}.
\end{barticle}
\endbibitem

\bibitem[\protect\citeauthoryear{{Kliem} and
  {T{\"o}r{\"o}k}}{2006}]{2006PhRvL..96y5002K}
\begin{barticle}
\bauthor{\bsnm{{Kliem}}, \binits{B.}},
\bauthor{\bsnm{{T{\"o}r{\"o}k}}, \binits{T.}}:
\byear{2006},
\batitle{{Torus Instability}}.
\bjtitle{Physical Review Letters}
\bvolume{96}(\bissue{25}),
\bfpage{255002}.
doi:\doiurl{10.1103/PhysRevLett.96.255002}.
\end{barticle}
\endbibitem

\bibitem[\protect\citeauthoryear{{Klimchuk}}{2001}]{2001AGUGM.125..143K}
\begin{barticle}
\bauthor{\bsnm{{Klimchuk}}, \binits{J.A.}}:
\byear{2001},
\batitle{{Theory of Coronal Mass Ejections}}.
\bjtitle{Space Weather (Geophysical Monograph 125), ed.~P.~Song, H.~Singer,
  G.~Siscoe (Washington: Am.~Geophys.~Un.)}
\bvolume{125},
\bfpage{143}.
\end{barticle}
\endbibitem

\bibitem[\protect\citeauthoryear{{Kopp} and
  {Pneuman}}{1976}]{1976SoPh...50...85K}
\begin{barticle}
\bauthor{\bsnm{{Kopp}}, \binits{R.A.}},
\bauthor{\bsnm{{Pneuman}}, \binits{G.W.}}:
\byear{1976},
\batitle{{Magnetic reconnection in the corona and the loop prominence
  phenomenon}}.
\bjtitle{\solphys}
\bvolume{50},
\bfpage{85}\,--\,\blpage{98}.
doi:\doiurl{10.1007/BF00206193}.
\end{barticle}
\endbibitem

\bibitem[\protect\citeauthoryear{{Krall}
  \textit{et~al.}}{2001}]{2001ApJ...562.1045K}
\begin{barticle}
\bauthor{\bsnm{{Krall}}, \binits{J.}},
\bauthor{\bsnm{{Chen}}, \binits{J.}},
\bauthor{\bsnm{{Duffin}}, \binits{R.T.}},
\bauthor{\bsnm{{Howard}}, \binits{R.A.}},
\bauthor{\bsnm{{Thompson}}, \binits{B.J.}}:
\byear{2001},
\batitle{{Erupting Solar Magnetic Flux Ropes: Theory and Observation}}.
\bjtitle{\apj}
\bvolume{562},
\bfpage{1045}\,--\,\blpage{1057}.
doi:\doiurl{10.1086/323844}.
\end{barticle}
\endbibitem

\bibitem[\protect\citeauthoryear{{Lemen}
  \textit{et~al.}}{2012}]{2012SoPh..275...17L}
\begin{barticle}
\bauthor{\bsnm{{Lemen}}, \binits{J.R.}},
\bauthor{\bsnm{{Title}}, \binits{A.M.}},
\bauthor{\bsnm{{Akin}}, \binits{D.J.}},
\bauthor{\bsnm{{Boerner}}, \binits{P.F.}},
\bauthor{\bsnm{{Chou}}, \binits{C.}},
\bauthor{\bsnm{{Drake}}, \binits{J.F.}},
\bauthor{\bsnm{{Duncan}}, \binits{D.W.}},
\bauthor{\bsnm{{Edwards}}, \binits{C.G.}},
\bauthor{\bsnm{{Friedlaender}}, \binits{F.M.}},
\bauthor{\bsnm{{Heyman}}, \binits{G.F.}},
\bauthor{\bsnm{{Hurlburt}}, \binits{N.E.}},
\bauthor{\bsnm{{Katz}}, \binits{N.L.}},
\bauthor{\bsnm{{Kushner}}, \binits{G.D.}},
\bauthor{\bsnm{{Levay}}, \binits{M.}},
\bauthor{\bsnm{{Lindgren}}, \binits{R.W.}},
\bauthor{\bsnm{{Mathur}}, \binits{D.P.}},
\bauthor{\bsnm{{McFeaters}}, \binits{E.L.}},
\bauthor{\bsnm{{Mitchell}}, \binits{S.}},
\bauthor{\bsnm{{Rehse}}, \binits{R.A.}},
\bauthor{\bsnm{{Schrijver}}, \binits{C.J.}},
\bauthor{\bsnm{{Springer}}, \binits{L.A.}},
\bauthor{\bsnm{{Stern}}, \binits{R.A.}},
\bauthor{\bsnm{{Tarbell}}, \binits{T.D.}},
\bauthor{\bsnm{{Wuelser}}, \binits{J.-P.}},
\bauthor{\bsnm{{Wolfson}}, \binits{C.J.}},
\bauthor{\bsnm{{Yanari}}, \binits{C.}},
\bauthor{\bsnm{{Bookbinder}}, \binits{J.A.}},
\bauthor{\bsnm{{Cheimets}}, \binits{P.N.}},
\bauthor{\bsnm{{Caldwell}}, \binits{D.}},
\bauthor{\bsnm{{Deluca}}, \binits{E.E.}},
\bauthor{\bsnm{{Gates}}, \binits{R.}},
\bauthor{\bsnm{{Golub}}, \binits{L.}},
\bauthor{\bsnm{{Park}}, \binits{S.}},
\bauthor{\bsnm{{Podgorski}}, \binits{W.A.}},
\bauthor{\bsnm{{Bush}}, \binits{R.I.}},
\bauthor{\bsnm{{Scherrer}}, \binits{P.H.}},
\bauthor{\bsnm{{Gummin}}, \binits{M.A.}},
\bauthor{\bsnm{{Smith}}, \binits{P.}},
\bauthor{\bsnm{{Auker}}, \binits{G.}},
\bauthor{\bsnm{{Jerram}}, \binits{P.}},
\bauthor{\bsnm{{Pool}}, \binits{P.}},
\bauthor{\bsnm{{Soufli}}, \binits{R.}},
\bauthor{\bsnm{{Windt}}, \binits{D.L.}},
\bauthor{\bsnm{{Beardsley}}, \binits{S.}},
\bauthor{\bsnm{{Clapp}}, \binits{M.}},
\bauthor{\bsnm{{Lang}}, \binits{J.}},
\bauthor{\bsnm{{Waltham}}, \binits{N.}}:
\byear{2012},
\batitle{{The Atmospheric Imaging Assembly (AIA) on the Solar Dynamics
  Observatory (SDO)}}.
\bjtitle{\solphys}
\bvolume{275},
\bfpage{17}\,--\,\blpage{40}.
doi:\doiurl{10.1007/s11207-011-9776-8}.
\end{barticle}
\endbibitem

\bibitem[\protect\citeauthoryear{{Lin}
  \textit{et~al.}}{2007}]{2007ApJ...658L.123L}
\begin{barticle}
\bauthor{\bsnm{{Lin}}, \binits{J.}},
\bauthor{\bsnm{{Li}}, \binits{J.}},
\bauthor{\bsnm{{Forbes}}, \binits{T.G.}},
\bauthor{\bsnm{{Ko}}, \binits{Y.}},
\bauthor{\bsnm{{Raymond}}, \binits{J.C.}},
\bauthor{\bsnm{{Vourlidas}}, \binits{A.}}:
\byear{2007},
\batitle{{Features and Properties of Coronal Mass Ejection/Flare Current
  Sheets}}.
\bjtitle{\apjl}
\bvolume{658},
\bfpage{L123}\,--\,\blpage{L126}.
doi:\doiurl{10.1086/515568}.
\end{barticle}
\endbibitem

\bibitem[\protect\citeauthoryear{{Liu}
  \textit{et~al.}}{2014}]{2014NatCo...5E3481L}
\begin{barticle}
\bauthor{\bsnm{{Liu}}, \binits{Y.D.}},
\bauthor{\bsnm{{Luhmann}}, \binits{J.G.}},
\bauthor{\bsnm{{Kajdi{\v c}}}, \binits{P.}},
\bauthor{\bsnm{{Kilpua}}, \binits{E.K.J.}},
\bauthor{\bsnm{{Lugaz}}, \binits{N.}},
\bauthor{\bsnm{{Nitta}}, \binits{N.V.}},
\bauthor{\bsnm{{M{\"o}stl}}, \binits{C.}},
\bauthor{\bsnm{{Lavraud}}, \binits{B.}},
\bauthor{\bsnm{{Bale}}, \binits{S.D.}},
\bauthor{\bsnm{{Farrugia}}, \binits{C.J.}},
\bauthor{\bsnm{{Galvin}}, \binits{A.B.}}:
\byear{2014},
\batitle{{Observations of an extreme storm in interplanetary space caused by
  successive coronal mass ejections}}.
\bjtitle{Nature Communications}
\bvolume{5}.
doi:\doiurl{10.1038/ncomms4481}.
\end{barticle}
\endbibitem

\bibitem[\protect\citeauthoryear{{Lockwood} and
  {Hapgood}}{2007}]{2007A&G....48f..11L}
\begin{barticle}
\bauthor{\bsnm{{Lockwood}}, \binits{M.}},
\bauthor{\bsnm{{Hapgood}}, \binits{M.}}:
\byear{2007},
\batitle{{The Rough Guide to the Moon and Mars}}.
\bjtitle{Astronomy and Geophysics}
\bvolume{48}(\bissue{6}),
\bfpage{060000}\,--\,\blpage{6}.
doi:\doiurl{10.1111/j.1468-4004.2007.48611.x}.
\end{barticle}
\endbibitem

\bibitem[\protect\citeauthoryear{{Lynch}
  \textit{et~al.}}{2008}]{2008ApJ...683.1192L}
\begin{barticle}
\bauthor{\bsnm{{Lynch}}, \binits{B.J.}},
\bauthor{\bsnm{{Antiochos}}, \binits{S.K.}},
\bauthor{\bsnm{{DeVore}}, \binits{C.R.}},
\bauthor{\bsnm{{Luhmann}}, \binits{J.G.}},
\bauthor{\bsnm{{Zurbuchen}}, \binits{T.H.}}:
\byear{2008},
\batitle{{Topological Evolution of a Fast Magnetic Breakout CME in Three
  Dimensions}}.
\bjtitle{\apj}
\bvolume{683},
\bfpage{1192}\,--\,\blpage{1206}.
doi:\doiurl{10.1086/589738}.
\end{barticle}
\endbibitem

\bibitem[\protect\citeauthoryear{{Moore} and
  {Labonte}}{1980}]{1980IAUS...91..207M}
\begin{bchapter}
\bauthor{\bsnm{{Moore}}, \binits{R.L.}},
\bauthor{\bsnm{{Labonte}}, \binits{B.J.}}:
\byear{1980},
\bctitle{{The filament eruption in the 3B flare of July 29, 1973 - Onset and
  magnetic field configuration}}.
In: \beditor{\bsnm{{Dryer}}, \binits{M.}},
\beditor{\bsnm{{Tandberg-Hanssen}}, \binits{E.}} (eds.)
\bbtitle{Solar and Interplanetary Dynamics},
\bsertitle{IAU Symposium}
\bseriesno{91},
\bfpage{207}\,--\,\blpage{210}.
\end{bchapter}
\endbibitem

\bibitem[\protect\citeauthoryear{{Morgan} and
  {Druckm{\"u}ller}}{2014}]{MorganDruckmuller_inreview}
\begin{botherref}
\oauthor{\bsnm{{Morgan}}, \binits{H.}},
\oauthor{\bsnm{{Druckm{\"u}ller}}, \binits{M.}}:
2014,
{Multi-scale Gaussian normalization for image processing: application to solar
  EUV observations}.
\textit{\solphys \,(in review)}.
\end{botherref}
\endbibitem

\bibitem[\protect\citeauthoryear{{Morgan}, {Byrne}, and
  {Habbal}}{2012}]{2012ApJ...752..144M}
\begin{barticle}
\bauthor{\bsnm{{Morgan}}, \binits{H.}},
\bauthor{\bsnm{{Byrne}}, \binits{J.P.}},
\bauthor{\bsnm{{Habbal}}, \binits{S.R.}}:
\byear{2012},
\batitle{{Automatically Detecting and Tracking Coronal Mass Ejections. I.
  Separation of Dynamic and Quiescent Components in Coronagraph Images}}.
\bjtitle{\apj}
\bvolume{752},
\bfpage{144}.
doi:\doiurl{10.1088/0004-637X/752/2/144}.
\end{barticle}
\endbibitem

\bibitem[\protect\citeauthoryear{{Morgan}, {Habbal}, and
  {Woo}}{2006}]{2006SoPh..236..263M}
\begin{barticle}
\bauthor{\bsnm{{Morgan}}, \binits{H.}},
\bauthor{\bsnm{{Habbal}}, \binits{S.R.}},
\bauthor{\bsnm{{Woo}}, \binits{R.}}:
\byear{2006},
\batitle{{The Depiction of Coronal Structure in White-Light Images}}.
\bjtitle{\solphys}
\bvolume{236},
\bfpage{263}\,--\,\blpage{272}.
doi:\doiurl{10.1007/s11207-006-0113-6}.
\end{barticle}
\endbibitem

\bibitem[\protect\citeauthoryear{{Morgan}, {Jeska}, and
  {Leonard}}{2013}]{2013ApJS..206...19M}
\begin{barticle}
\bauthor{\bsnm{{Morgan}}, \binits{H.}},
\bauthor{\bsnm{{Jeska}}, \binits{L.}},
\bauthor{\bsnm{{Leonard}}, \binits{D.}}:
\byear{2013},
\batitle{{The Expansion of Active Regions into the Extended Solar Corona}}.
\bjtitle{\apjs}
\bvolume{206},
\bfpage{19}.
doi:\doiurl{10.1088/0067-0049/206/2/19}.
\end{barticle}
\endbibitem

\bibitem[\protect\citeauthoryear{{P\'erez-Su\'arez}
  \textit{et~al.}}{2011}]{2011igi-global}
\begin{bbook}
\bauthor{\bsnm{{P\'erez-Su\'arez}}, \binits{D.}},
\bauthor{\bsnm{{Higgins}}, \binits{P.A.}},
\bauthor{\bsnm{{Bloomfield}}, \binits{D.S.}},
\bauthor{\bsnm{{McAteer}}, \binits{R.T.J.}},
\bauthor{\bsnm{{Krista}}, \binits{L.D.}},
\bauthor{\bsnm{{Byrne}}, \binits{J.P.}},
\bauthor{\bsnm{{Gallagher}}, \binits{P.T.}}:
\byear{2011},
\bbtitle{{``Automated Solar Feature Detection for Space Weather Applications",
  in Applied Signal and Image Processing: Multidisciplinary Advancements, eds.
  R. Qahwaji, R. Green, \& E. L. Hines, (IGI Global), p. 207\,--\,225}}.
\end{bbook}
\endbibitem

\bibitem[\protect\citeauthoryear{{Pesnell}, {Thompson}, and
  {Chamberlin}}{2012}]{2012SoPh..275....3P}
\begin{barticle}
\bauthor{\bsnm{{Pesnell}}, \binits{W.D.}},
\bauthor{\bsnm{{Thompson}}, \binits{B.J.}},
\bauthor{\bsnm{{Chamberlin}}, \binits{P.C.}}:
\byear{2012},
\batitle{{The Solar Dynamics Observatory (SDO)}}.
\bjtitle{\solphys}
\bvolume{275},
\bfpage{3}\,--\,\blpage{15}.
doi:\doiurl{10.1007/s11207-011-9841-3}.
\end{barticle}
\endbibitem

\bibitem[\protect\citeauthoryear{{Prang{\'e}}
  \textit{et~al.}}{2004}]{2004Natur.432...78P}
\begin{barticle}
\bauthor{\bsnm{{Prang{\'e}}}, \binits{R.}},
\bauthor{\bsnm{{Pallier}}, \binits{L.}},
\bauthor{\bsnm{{Hansen}}, \binits{K.C.}},
\bauthor{\bsnm{{Howard}}, \binits{R.}},
\bauthor{\bsnm{{Vourlidas}}, \binits{A.}},
\bauthor{\bsnm{{Courtin}}, \binits{R.}},
\bauthor{\bsnm{{Parkinson}}, \binits{C.}}:
\byear{2004},
\batitle{{An interplanetary shock traced by planetary auroral storms from the
  Sun to Saturn}}.
\bjtitle{\nat}
\bvolume{432},
\bfpage{78}\,--\,\blpage{81}.
doi:\doiurl{10.1038/nature02986}.
\end{barticle}
\endbibitem

\bibitem[\protect\citeauthoryear{{Priest} and
  {Forbes}}{2002}]{2002A&ARv..10..313P}
\begin{barticle}
\bauthor{\bsnm{{Priest}}, \binits{E.R.}},
\bauthor{\bsnm{{Forbes}}, \binits{T.G.}}:
\byear{2002},
\batitle{{The magnetic nature of solar flares}}.
\bjtitle{\aapr}
\bvolume{10},
\bfpage{313}\,--\,\blpage{377}.
doi:\doiurl{10.1007/s001590100013}.
\end{barticle}
\endbibitem

\bibitem[\protect\citeauthoryear{{Raftery}
  \textit{et~al.}}{2010}]{2010ApJ...721.1579R}
\begin{barticle}
\bauthor{\bsnm{{Raftery}}, \binits{C.L.}},
\bauthor{\bsnm{{Gallagher}}, \binits{P.T.}},
\bauthor{\bsnm{{McAteer}}, \binits{R.T.J.}},
\bauthor{\bsnm{{Lin}}, \binits{C.-H.}},
\bauthor{\bsnm{{Delahunt}}, \binits{G.}}:
\byear{2010},
\batitle{{Evidence for Internal Tether-cutting in a Flare/Coronal Mass Ejection
  Observed by MESSENGER, RHESSI, and STEREO}}.
\bjtitle{\apj}
\bvolume{721},
\bfpage{1579}\,--\,\blpage{1584}.
doi:\doiurl{10.1088/0004-637X/721/2/1579}.
\end{barticle}
\endbibitem

\bibitem[\protect\citeauthoryear{{Santandrea}
  \textit{et~al.}}{2013}]{2013SoPh..286....5S}
\begin{barticle}
\bauthor{\bsnm{{Santandrea}}, \binits{S.}},
\bauthor{\bsnm{{Gantois}}, \binits{K.}},
\bauthor{\bsnm{{Strauch}}, \binits{K.}},
\bauthor{\bsnm{{Teston}}, \binits{F.}},
\bauthor{\bsnm{{Tilmans}}, \binits{E.}},
\bauthor{\bsnm{{Baijot}}, \binits{C.}},
\bauthor{\bsnm{{Gerrits}}, \binits{D.}},
\bauthor{\bsnm{{De Groof}}, \binits{A.}},
\bauthor{\bsnm{{Schwehm}}, \binits{G.}},
\bauthor{\bsnm{{Zender}}, \binits{J.}}:
\byear{2013},
\batitle{{PROBA2: Mission and Spacecraft Overview}}.
\bjtitle{\solphys}
\bvolume{286},
\bfpage{5}\,--\,\blpage{19}.
doi:\doiurl{10.1007/s11207-013-0289-5}.
\end{barticle}
\endbibitem

\bibitem[\protect\citeauthoryear{{Schrijver}
  \textit{et~al.}}{2008}]{2008ApJ...674..586S}
\begin{barticle}
\bauthor{\bsnm{{Schrijver}}, \binits{C.J.}},
\bauthor{\bsnm{{Elmore}}, \binits{C.}},
\bauthor{\bsnm{{Kliem}}, \binits{B.}},
\bauthor{\bsnm{{T{\"o}r{\"o}k}}, \binits{T.}},
\bauthor{\bsnm{{Title}}, \binits{A.M.}}:
\byear{2008},
\batitle{{Observations and Modeling of the Early Acceleration Phase of Erupting
  Filaments Involved in Coronal Mass Ejections}}.
\bjtitle{\apj}
\bvolume{674},
\bfpage{586}\,--\,\blpage{595}.
doi:\doiurl{10.1086/524294}.
\end{barticle}
\endbibitem

\bibitem[\protect\citeauthoryear{{Schwenn}
  \textit{et~al.}}{2005}]{2005AnGeo..23.1033S}
\begin{barticle}
\bauthor{\bsnm{{Schwenn}}, \binits{R.}},
\bauthor{\bsnm{{dal Lago}}, \binits{A.}},
\bauthor{\bsnm{{Huttunen}}, \binits{E.}},
\bauthor{\bsnm{{Gonzalez}}, \binits{W.D.}}:
\byear{2005},
\batitle{{The association of coronal mass ejections with their effects near the
  Earth}}.
\bjtitle{Annales Geophysicae}
\bvolume{23},
\bfpage{1033}\,--\,\blpage{1059}.
\end{barticle}
\endbibitem

\bibitem[\protect\citeauthoryear{{Seaton}
  \textit{et~al.}}{2013}]{2013SoPh..286...43S}
\begin{barticle}
\bauthor{\bsnm{{Seaton}}, \binits{D.B.}},
\bauthor{\bsnm{{Berghmans}}, \binits{D.}},
\bauthor{\bsnm{{Nicula}}, \binits{B.}},
\bauthor{\bsnm{{Halain}}, \binits{J.-P.}},
\bauthor{\bsnm{{De Groof}}, \binits{A.}},
\bauthor{\bsnm{{Thibert}}, \binits{T.}},
\bauthor{\bsnm{{Bloomfield}}, \binits{D.S.}},
\bauthor{\bsnm{{Raftery}}, \binits{C.L.}},
\bauthor{\bsnm{{Gallagher}}, \binits{P.T.}},
\bauthor{\bsnm{{Auch{\`e}re}}, \binits{F.}},
\bauthor{\bsnm{{Defise}}, \binits{J.-M.}},
\bauthor{\bsnm{{D'Huys}}, \binits{E.}},
\bauthor{\bsnm{{Lecat}}, \binits{J.-H.}},
\bauthor{\bsnm{{Mazy}}, \binits{E.}},
\bauthor{\bsnm{{Rochus}}, \binits{P.}},
\bauthor{\bsnm{{Rossi}}, \binits{L.}},
\bauthor{\bsnm{{Sch{\"u}hle}}, \binits{U.}},
\bauthor{\bsnm{{Slemzin}}, \binits{V.}},
\bauthor{\bsnm{{Yalim}}, \binits{M.S.}},
\bauthor{\bsnm{{Zender}}, \binits{J.}}:
\byear{2013},
\batitle{{The SWAP EUV Imaging Telescope Part I: Instrument Overview and
  Pre-Flight Testing}}.
\bjtitle{\solphys}
\bvolume{286},
\bfpage{43}\,--\,\blpage{65}.
doi:\doiurl{10.1007/s11207-012-0114-6}.
\end{barticle}
\endbibitem

\bibitem[\protect\citeauthoryear{{Stenborg} and
  {Cobelli}}{2003}]{2003A&A...398.1185S}
\begin{barticle}
\bauthor{\bsnm{{Stenborg}}, \binits{G.}},
\bauthor{\bsnm{{Cobelli}}, \binits{P.J.}}:
\byear{2003},
\batitle{{A wavelet packets equalization technique to reveal the multiple
  spatial-scale nature of coronal structures}}.
\bjtitle{\aap}
\bvolume{398},
\bfpage{1185}\,--\,\blpage{1193}.
doi:\doiurl{10.1051/0004-6361:20021687}.
\end{barticle}
\endbibitem

\bibitem[\protect\citeauthoryear{{Stenborg}, {Vourlidas}, and
  {Howard}}{2008}]{2008ApJ...674.1201S}
\begin{barticle}
\bauthor{\bsnm{{Stenborg}}, \binits{G.}},
\bauthor{\bsnm{{Vourlidas}}, \binits{A.}},
\bauthor{\bsnm{{Howard}}, \binits{R.A.}}:
\byear{2008},
\batitle{{A Fresh View of the Extreme-Ultraviolet Corona from the Application
  of a New Image-Processing Technique}}.
\bjtitle{\apj}
\bvolume{674},
\bfpage{1201}\,--\,\blpage{1206}.
doi:\doiurl{10.1086/525556}.
\end{barticle}
\endbibitem

\bibitem[\protect\citeauthoryear{{Sturrock}}{1966}]{1966Natur.211..695S}
\begin{barticle}
\bauthor{\bsnm{{Sturrock}}, \binits{P.A.}}:
\byear{1966},
\batitle{{Model of the High-Energy Phase of Solar Flares}}.
\bjtitle{\nat}
\bvolume{211},
\bfpage{695}\,--\,\blpage{697}.
doi:\doiurl{10.1038/211695a0}.
\end{barticle}
\endbibitem

\bibitem[\protect\citeauthoryear{{Su}
  \textit{et~al.}}{2012}]{2012ApJ...746L...5S}
\begin{barticle}
\bauthor{\bsnm{{Su}}, \binits{Y.}},
\bauthor{\bsnm{{Dennis}}, \binits{B.R.}},
\bauthor{\bsnm{{Holman}}, \binits{G.D.}},
\bauthor{\bsnm{{Wang}}, \binits{T.}},
\bauthor{\bsnm{{Chamberlin}}, \binits{P.C.}},
\bauthor{\bsnm{{Savage}}, \binits{S.}},
\bauthor{\bsnm{{Veronig}}, \binits{A.}}:
\byear{2012},
\batitle{{Observations of a Two-stage Solar Eruptive Event (SEE): Evidence for
  Secondary Heating}}.
\bjtitle{\apjl}
\bvolume{746},
\bfpage{L5}.
doi:\doiurl{10.1088/2041-8205/746/1/L5}.
\end{barticle}
\endbibitem

\bibitem[\protect\citeauthoryear{{Subramanian} and
  {Dere}}{2001}]{2001ApJ...561..372S}
\begin{barticle}
\bauthor{\bsnm{{Subramanian}}, \binits{P.}},
\bauthor{\bsnm{{Dere}}, \binits{K.P.}}:
\byear{2001},
\batitle{{Source Regions of Coronal Mass Ejections}}.
\bjtitle{\apj}
\bvolume{561},
\bfpage{372}\,--\,\blpage{395}.
doi:\doiurl{10.1086/323213}.
\end{barticle}
\endbibitem

\bibitem[\protect\citeauthoryear{{Titov} and
  {D{\'e}moulin}}{1999}]{1999A&A...351..707T}
\begin{barticle}
\bauthor{\bsnm{{Titov}}, \binits{V.S.}},
\bauthor{\bsnm{{D{\'e}moulin}}, \binits{P.}}:
\byear{1999},
\batitle{{Basic topology of twisted magnetic configurations in solar flares}}.
\bjtitle{\aap}
\bvolume{351},
\bfpage{707}\,--\,\blpage{720}.
\end{barticle}
\endbibitem

\bibitem[\protect\citeauthoryear{{T{\"o}r{\"o}k}, {Kliem}, and
  {Titov}}{2004}]{2004A&A...413L..27T}
\begin{barticle}
\bauthor{\bsnm{{T{\"o}r{\"o}k}}, \binits{T.}},
\bauthor{\bsnm{{Kliem}}, \binits{B.}},
\bauthor{\bsnm{{Titov}}, \binits{V.S.}}:
\byear{2004},
\batitle{{Ideal kink instability of a magnetic loop equilibrium}}.
\bjtitle{\aap}
\bvolume{413},
\bfpage{L27}\,--\,\blpage{L30}.
doi:\doiurl{10.1051/0004-6361:20031691}.
\end{barticle}
\endbibitem

\bibitem[\protect\citeauthoryear{{van der Holst}, {Jacobs}, and
  {Poedts}}{2007}]{2007ApJ...671L..77V}
\begin{barticle}
\bauthor{\bsnm{{van der Holst}}, \binits{B.}},
\bauthor{\bsnm{{Jacobs}}, \binits{C.}},
\bauthor{\bsnm{{Poedts}}, \binits{S.}}:
\byear{2007},
\batitle{{Simulation of a Breakout Coronal Mass Ejection in the Solar Wind}}.
\bjtitle{\apjl}
\bvolume{671},
\bfpage{L77}\,--\,\blpage{L80}.
doi:\doiurl{10.1086/524732}.
\end{barticle}
\endbibitem

\bibitem[\protect\citeauthoryear{{van der Holst}
  \textit{et~al.}}{2009}]{2009ApJ...693.1178V}
\begin{barticle}
\bauthor{\bsnm{{van der Holst}}, \binits{B.}},
\bauthor{\bsnm{{Manchester}}, \binits{W.} \bsuffix{IV}},
\bauthor{\bsnm{{Sokolov}}, \binits{I.V.}},
\bauthor{\bsnm{{T{\'o}th}}, \binits{G.}},
\bauthor{\bsnm{{Gombosi}}, \binits{T.I.}},
\bauthor{\bsnm{{DeZeeuw}}, \binits{D.}},
\bauthor{\bsnm{{Cohen}}, \binits{O.}}:
\byear{2009},
\batitle{{Breakout Coronal Mass Ejection or Streamer Blowout: The Bugle
  Effect}}.
\bjtitle{\apj}
\bvolume{693},
\bfpage{1178}\,--\,\blpage{1187}.
doi:\doiurl{10.1088/0004-637X/693/2/1178}.
\end{barticle}
\endbibitem

\bibitem[\protect\citeauthoryear{{Wang}
  \textit{et~al.}}{2012}]{2012ApJ...749..182W}
\begin{barticle}
\bauthor{\bsnm{{Wang}}, \binits{Y.-M.}},
\bauthor{\bsnm{{Grappin}}, \binits{R.}},
\bauthor{\bsnm{{Robbrecht}}, \binits{E.}},
\bauthor{\bsnm{{Sheeley}}, \binits{N.R.} \bsuffix{Jr.}}:
\byear{2012},
\batitle{{On the Nature of the Solar Wind from Coronal Pseudostreamers}}.
\bjtitle{\apj}
\bvolume{749},
\bfpage{182}.
doi:\doiurl{10.1088/0004-637X/749/2/182}.
\end{barticle}
\endbibitem

\bibitem[\protect\citeauthoryear{{Webb} and
  {Howard}}{2012}]{2012LRSP....9....3W}
\begin{barticle}
\bauthor{\bsnm{{Webb}}, \binits{D.F.}},
\bauthor{\bsnm{{Howard}}, \binits{T.A.}}:
\byear{2012},
\batitle{{Coronal Mass Ejections: Observations}}.
\bjtitle{Living Reviews in Solar Physics}
\bvolume{9},
\bfpage{3}.
\end{barticle}
\endbibitem

\bibitem[\protect\citeauthoryear{{Wuelser}
  \textit{et~al.}}{2004}]{2004SPIE.5171..111W}
\begin{bchapter}
\bauthor{\bsnm{{Wuelser}}, \binits{J.}},
\bauthor{\bsnm{{Lemen}}, \binits{J.R.}},
\bauthor{\bsnm{{Tarbell}}, \binits{T.D.}},
\bauthor{\bsnm{{Wolfson}}, \binits{C.J.}},
\bauthor{\bsnm{{Cannon}}, \binits{J.C.}},
\bauthor{\bsnm{{Carpenter}}, \binits{B.A.}},
\bauthor{\bsnm{{Duncan}}, \binits{D.W.}},
\bauthor{\bsnm{{Gradwohl}}, \binits{G.S.}},
\bauthor{\bsnm{{Meyer}}, \binits{S.B.}},
\bauthor{\bsnm{{Moore}}, \binits{A.S.}},
\bauthor{\bsnm{{Navarro}}, \binits{R.L.}},
\bauthor{\bsnm{{Pearson}}, \binits{J.D.}},
\bauthor{\bsnm{{Rossi}}, \binits{G.R.}},
\bauthor{\bsnm{{Springer}}, \binits{L.A.}},
\bauthor{\bsnm{{Howard}}, \binits{R.A.}},
\bauthor{\bsnm{{Moses}}, \binits{J.D.}},
\bauthor{\bsnm{{Newmark}}, \binits{J.S.}},
\bauthor{\bsnm{{Delaboudiniere}}, \binits{J.}},
\bauthor{\bsnm{{Artzner}}, \binits{G.E.}},
\bauthor{\bsnm{{Auchere}}, \binits{F.}},
\bauthor{\bsnm{{Bougnet}}, \binits{M.}},
\bauthor{\bsnm{{Bouyries}}, \binits{P.}},
\bauthor{\bsnm{{Bridou}}, \binits{F.}},
\bauthor{\bsnm{{Clotaire}}, \binits{J.}},
\bauthor{\bsnm{{Colas}}, \binits{G.}},
\bauthor{\bsnm{{Delmotte}}, \binits{F.}},
\bauthor{\bsnm{{Jerome}}, \binits{A.}},
\bauthor{\bsnm{{Lamare}}, \binits{M.}},
\bauthor{\bsnm{{Mercier}}, \binits{R.}},
\bauthor{\bsnm{{Mullot}}, \binits{M.}},
\bauthor{\bsnm{{Ravet}}, \binits{M.}},
\bauthor{\bsnm{{Song}}, \binits{X.}},
\bauthor{\bsnm{{Bothmer}}, \binits{V.}},
\bauthor{\bsnm{{Deutsch}}, \binits{W.}}:
\byear{2004},
\bctitle{{EUVI: the STEREO-SECCHI extreme ultraviolet imager}}.
In: \bbtitle{\emph{Fineschi, S. \& Gummin, M.~A., ed.,} Society of
  Photo-Optical Instrumentation Engineers (SPIE) Conference Series, \emph{{\bf
  5171}, 111-122}}.
\end{bchapter}
\endbibitem

\bibitem[\protect\citeauthoryear{{Yashiro}
  \textit{et~al.}}{2004}]{2004JGRA..10907105Y}
\begin{barticle}
\bauthor{\bsnm{{Yashiro}}, \binits{S.}},
\bauthor{\bsnm{{Gopalswamy}}, \binits{N.}},
\bauthor{\bsnm{{Michalek}}, \binits{G.}},
\bauthor{\bsnm{{St.~Cyr}}, \binits{O.C.}},
\bauthor{\bsnm{{Plunkett}}, \binits{S.P.}},
\bauthor{\bsnm{{Rich}}, \binits{N.B.}},
\bauthor{\bsnm{{Howard}}, \binits{R.A.}}:
\byear{2004},
\batitle{{A catalog of white light coronal mass ejections observed by the SOHO
  spacecraft}}.
\bjtitle{Journal of Geophysical Research (Space Physics)}
\bvolume{109},
\bfpage{7105}.
doi:\doiurl{10.1029/2003JA010282}.
\end{barticle}
\endbibitem

\bibitem[\protect\citeauthoryear{{Young} and
  {Gallagher}}{2008}]{2008SoPh..248..457Y}
\begin{barticle}
\bauthor{\bsnm{{Young}}, \binits{C.A.}},
\bauthor{\bsnm{{Gallagher}}, \binits{P.T.}}:
\byear{2008},
\batitle{{Multiscale Edge Detection in the Corona}}.
\bjtitle{\solphys}
\bvolume{248},
\bfpage{457}\,--\,\blpage{469}.
doi:\doiurl{10.1007/s11207-008-9177-9}.
\end{barticle}
\endbibitem

\bibitem[\protect\citeauthoryear{{Zhang} and
  {Wang}}{2002}]{2002ApJ...566L.117Z}
\begin{barticle}
\bauthor{\bsnm{{Zhang}}, \binits{J.}},
\bauthor{\bsnm{{Wang}}, \binits{J.}}:
\byear{2002},
\batitle{{Are Homologous Flare-Coronal Mass Ejection Events Triggered by Moving
  Magnetic Features?}}
\bjtitle{\apj}
\bvolume{566},
\bfpage{L117}\,--\,\blpage{L120}.
\burl{http://adsabs.harvard.edu/abs/2002ApJ...566L.117Z}.
\end{barticle}
\endbibitem

\bibitem[\protect\citeauthoryear{{Zhou}
  \textit{et~al.}}{2006}]{2006ApJ...651.1238Z}
\begin{barticle}
\bauthor{\bsnm{{Zhou}}, \binits{G.P.}},
\bauthor{\bsnm{{Wang}}, \binits{J.X.}},
\bauthor{\bsnm{{Zhang}}, \binits{J.}},
\bauthor{\bsnm{{Chen}}, \binits{P.F.}},
\bauthor{\bsnm{{Ji}}, \binits{H.S.}},
\bauthor{\bsnm{{Dere}}, \binits{K.}}:
\byear{2006},
\batitle{{Two Successive Coronal Mass Ejections Driven by the Kink and Drainage
  Instabilities of an Eruptive Prominence}}.
\bjtitle{\apj}
\bvolume{651},
\bfpage{1238}\,--\,\blpage{1244}.
doi:\doiurl{10.1086/507977}.
\end{barticle}
\endbibitem

\bibitem[\protect\citeauthoryear{{Zuccarello}
  \textit{et~al.}}{2014}]{2014zuccarello}
\begin{barticle}
\bauthor{\bsnm{{Zuccarello}}, \binits{F.P.}},
\bauthor{\bsnm{{Seaton}}, \binits{D.B.}},
\bauthor{\bsnm{{Mierla}}, \binits{M.}},
\bauthor{\bsnm{{Poedts}}, \binits{S.}},
\bauthor{\bsnm{{Rachmeler}}, \binits{L.A.}},
\bauthor{\bsnm{{Romano}}, \binits{P.}},
\bauthor{\bsnm{{Zuccarello}}, \binits{F.}}:
\byear{2014},
\batitle{{Observational evidence of torus instability as trigger mechanism for
  coronal mass ejections: the 2011 August 4 filament eruption}}.
\bjtitle{\apj}
\bvolume{785},
\bfpage{88}.
doi:\doiurl{10.1088/0004-637X/785/2/88}.
\end{barticle}
\endbibitem

\end{thebibliography}
%
%
%
%

\end{article} 
\end{document}